\documentclass[aps, prd, 10pt, twocolumn, superscriptaddress,noshowpacs, preprintnumbers, longbibliography,nofootinbib,bibnotes,hyperref,floatfix]{revtex4-1}

\usepackage{amsmath}
\usepackage{amsfonts}
\usepackage{amssymb}
\usepackage{bbold}
\usepackage{epsfig}
\usepackage{graphicx}
\usepackage{bm}
\usepackage{array}
\usepackage{hyperref}
\usepackage{listings}
\usepackage{color}
\usepackage{float}
\usepackage{url} 
\usepackage[normalem]{ulem}

\usepackage{tikz,xcolor,hyperref}

\definecolor{lime}{HTML}{A6CE39}
\DeclareRobustCommand{\orcidicon}{\hspace{-1mm}
 \begin{tikzpicture}
 \draw[lime, fill=lime] (0,0) 
 circle [radius=0.16] 
 node[white] {{\fontfamily{qag}\selectfont \tiny \,ID}};
 \draw[white, fill=white] (-0.0525,0.095) 
 circle [radius=0.007];
 \end{tikzpicture}
 \hspace{-3mm}
}

\foreach \x in {A, ..., Z}{\expandafter\xdef\csname orcid\x\endcsname{\noexpand\href{https://orcid.org/\csname orcidauthor\x\endcsname}
 {\noexpand\orcidicon}}
}

\begin{document}

\title{Do Neutrinos Become Flavor Unstable Due to Collisions with Matter in the Supernova Decoupling Region?}

\author{Shashank Shalgar\orcidA{}}
\affiliation{Niels Bohr International Academy \& DARK, Niels Bohr Institute,\\University of Copenhagen, Blegdamsvej 17, 2100 Copenhagen, Denmark}
\author{Irene Tamborra\orcidB{}}
\affiliation{Niels Bohr International Academy \& DARK, Niels Bohr Institute,\\University of Copenhagen, Blegdamsvej 17, 2100 Copenhagen, Denmark}

\date{\today}

\begin{abstract}
In core-collapse supernovae, the neutrino density is so large that neutrino flavor instabilities, leading to flavor conversion, can be triggered by the forward scattering of neutrinos among each other, if a crossing between the angular distributions of electron neutrinos and antineutrinos exists (fast instability in the limit of vanishing vacuum frequency) or in the presence of perturbations induced by the neutrino vacuum frequency (slow instability). Recently, it has been advanced the conjecture that neutrino collisions with the medium could be another mean to kickstart  flavor change (collisional instability). Inspired by a spherically symmetric core-collapse supernova model with mass $18.6\ M_\odot$, we compute the neutrino angular distributions solving the kinetic equations for an average energy mode and investigate the occurrence of flavor instabilities at different post-bounce times, ranging from the accretion phase to the early cooling phase. We find that fast and slow flavor instabilities largely dominate over the collisional ones in the decoupling region for all post-bounce times. While more work is needed to assess the relevance of collisional instabilities in neutrino-dense  environments, our findings suggest that neutrino collisions with matter affect the flavor evolution in the decoupling region, but are not responsible for triggering flavor conversion, if crossings in the neutrino lepton number angular distribution exist. 
\end{abstract}

\maketitle

\section{Introduction}
\label{sec:intro}

In dense neutrino astrophysical environments, such as the interior of core-collapse supernovae (SNe) and neutron star mergers, neutrino flavor evolution has a rich phenomenology owing to the nonlinear nature of its evolution~\cite{Mirizzi:2015eza, Duan:2010bg, Chakraborty:2016yeg, Tamborra:2020cul,Richers:2022zug}. The nonlinearity is a consequence of the coherent forward scattering of neutrinos among each other. Flavor conversion is classified as fast, if purely driven by the neutrino-neutrino interaction term for vanishing neutrino vacuum frequency~\cite{Sawyer:2005jk, Sawyer:2008zs, Sawyer:2015dsa, Chakraborty:2016lct, Izaguirre:2016gsx, Tamborra:2020cul}; in this case, a necessary but not sufficient condition for the development of fast flavor instabilities is the existence of electron lepton number (ELN) crossings in the angular distribution of electron neutrinos and antineutrinos (note that, under the assumption of periodic boundary conditions, the existence of an ELN crossing becomes a sufficient condition for the development of a flavor instability), albeit a large growth rate for such flavor instabilities does not necessarily imply large flavor conversion~\cite{Padilla-Gay:2021haz,Morinaga:2021vmc}.  In addition, neutrino self-interactions could be triggered at high densities by the vacuum frequency in the Hamiltonian~\cite{Shalgar:2022rjj, Shalgar:2022lvv}; in this case, we consider the flavor instability to be slow since it develops thanks to the vacuum frequency, which is smaller than the neutrino-neutrino interaction strength; note that, in this case, a system with ELN crossings which is stable for vanishing vacuum frequency may become unstable. 

Owing to the nonlinearity of the system, the momentum changing collisional processes, which are lower order processes, can have a significant impact on the flavor evolution, despite the widely different characteristic time scales characterizing flavor evolution and collisions~\cite{Shalgar:2020wcx, Johns:2021qby, Sasaki:2021zld, Hansen:2022xza, Kato:2022vsu, Johns:2022yqy, Padilla-Gay:2022wck, Lin:2022dek, Liu:2023pjw, Kato:2023dcw}. 
The collisional processes of interest include momentum changing scattering as well as reactions responsible for the creation and absorption of neutrinos~\cite{Burrows:2004vq,Janka:2012wk,Mezzacappa:2020oyq,Shalgar:2019kzy}. As the neutrino emission properties are affected by the collisional processes, before and during flavor evolution, it is imperative to understand their interplay. First attempts coupling collisions and neutrino flavor evolution self-consistently have been reported in the literature, but there is still a long journey ahead~\cite{Shalgar:2022rjj, Shalgar:2022lvv, Nagakura:2023mhr,Xiong:2022vsy}.

Existing work employing a given initial ELN angular distribution with a crossing finds that the effect of collisions on flavor evolution is to isotropize the angular distribution, with a resulting enhancement or suppression of flavor conversion according to the initial ELN distributions~\cite{Shalgar:2020wcx, Sasaki:2021zld, Hansen:2022xza,Martin:2021xyl}. 
On the other hand, a system  with no flavor instability in the absence of the collision term can also become unstable in the flavor space because of the so-called collisional instabilities~\cite{Johns:2021qby, Johns:2022yqy, Martin:2021xyl, Lin:2022dek,Padilla-Gay:2022wck,Xiong:2022vsy,Xiong:2022zqz}. 

In this paper, we aim to investigate slow~\footnote{In this work, slow flavor instability indicates a flavor instability that vanishes for zero vacuum frequency, independently on the existence of crossings in the lepton number angular distribution. For fast instabilities, instead, we mean instabilities that arise in the limit of vanishing vacuum frequency.}, fast, and collisional flavor instabilities, not relying on arbitrary neutrino angular distributions, rather solving the neutrino kinetic equations with the collision terms for an average energy mode to obtain the neutrino angular distributions for each flavor. To this purpose, we rely on static hydrodynamic backgrounds and thermodynamic properties extrapolated at different post-bounce snapshots of a spherically symmetric core-collapse SN model with mass $18.6\ M_\odot$~\cite{SNarchive}. To compute the neutrino angular distributions, in the collisional kernel, we include emission and absorption of neutrinos through Bremsstrahlung, pair production, and beta reactions. In addition, we consider the direction changing effects due to neutral current scattering of neutrinos on nucleons. 
We then investigate through the linear stability analysis whether collisional, slow, or fast flavor instabilities exist.

The manuscript is organized as follows. Section~\ref{sec:qke} introduces the neutrino quantum kinetic equations, then we discuss the procedure adopted to obtain the neutrino angular distributions solving the kinetic equations in the absence of flavor conversion in Sec.~\ref{sec:steady}. In Sec.~\ref{sec:lsa}, we perform the linear stability analysis for the homogeneous mode, with as well as without the collision term in order to understand the effect of collisional instabilities, while we discuss the impact of spatial  inhomogeneities in Sec.~\ref{sec:lsa_inh}. We summarize our findings and offer an outlook in 
Sec.~\ref{sec:conclusions}. In addition, Appendix~\ref{collisions} provides an outline of the collision terms responsible for the production, absorption, and momentum changing interactions; while a comparison of the neutrino number densities computed in this work with the ones obtained from the hydrodynamical SN simulation is shown in Appendix~\ref{sec:comparison}.

\section{Neutrino quantum kinetic equations}
\label{sec:qke}
\begin{figure}
\includegraphics[width=0.49\textwidth]{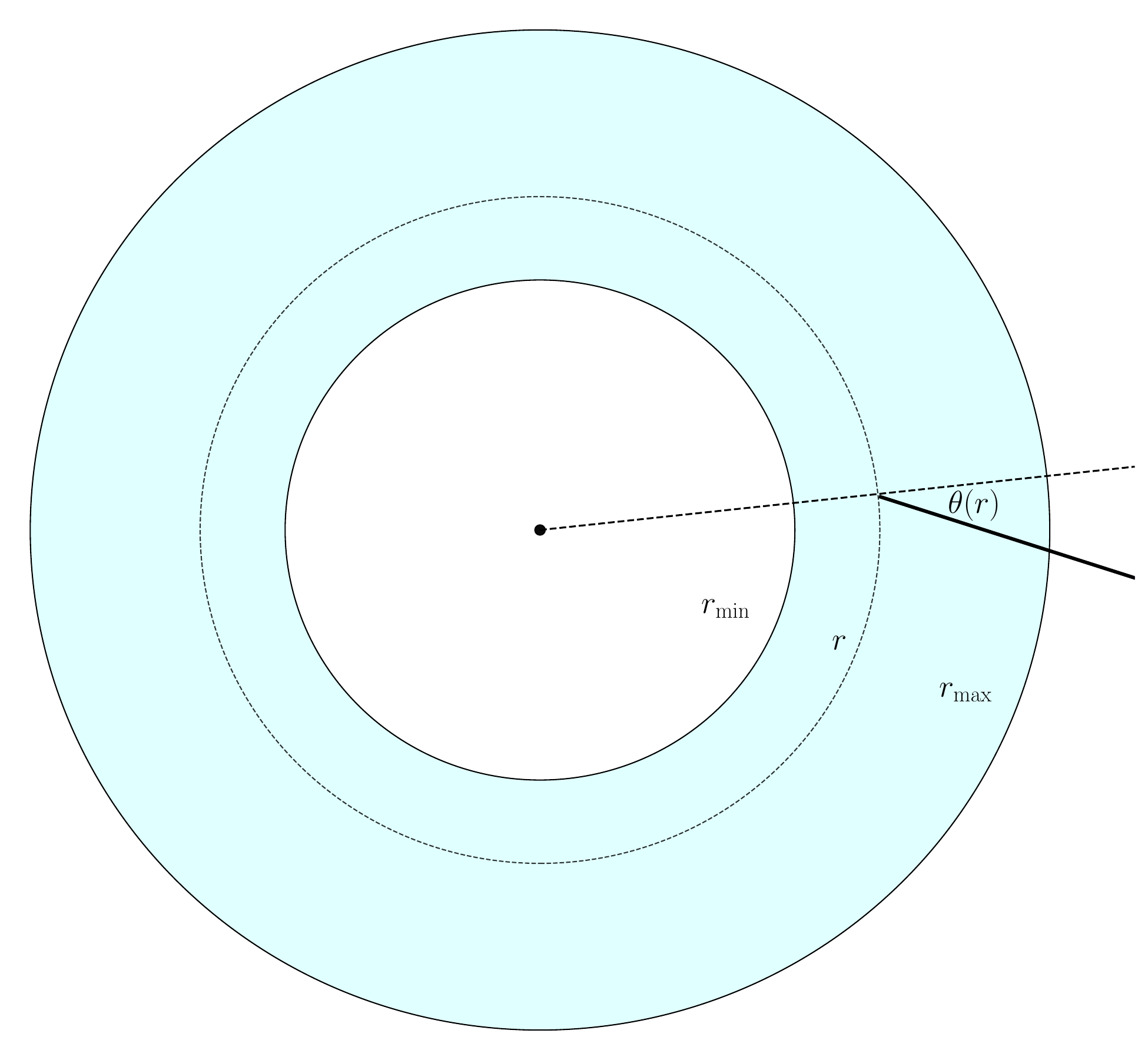}
\caption{Schematic diagram of the shell within which we solve the neutrino kinetic equations taking into account emission, absorption, and direction changing processes. The polar angle $\theta$ is defined with respect to the radial direction. It should be noted that $\theta$ changes with the radius for a given trajectory.}
\label{cartoon}
\end{figure}

For the sake of simplicity, we consider two monoenergetic neutrino flavors, $\nu_e$ and $\nu_x$ with $x=\mu, \tau$ (or a linear combination of the two).
The equations of motion of neutrinos and antineutrinos are 
\begin{eqnarray}
i\left(\frac{\partial }{\partial t} + \vec{v}\cdot \vec{\nabla}\right)\rho &=& [H,\rho] + i\mathcal{C}[\rho,\bar{\rho}]\ ,%\mathcal{C}_{\textrm{emit}} - \mathcal{C}_{\textrm{absorb}}\rho + C_{\textrm{dir-ch}}\ ,
\label{vlasov1}
\\
i\left(\frac{\partial }{\partial t} + \vec{v}\cdot \vec{\nabla}\right)\bar{\rho} &=& [\bar{H},\bar\rho] + i\bar{\mathcal{C}}[\rho, \bar{\rho}]\ , %\mathcal{C}_{\textrm{emit}} - \mathcal{C}_{\textrm{absorb}}\bar{\rho} + C_{\textrm{dir-ch}}\ .
\label{vlasov2}
\end{eqnarray}
with $\rho$ and $\bar{\rho}$ being $2\times 2$ density matrices representing the flavor content of neutrinos and antineutrinos, respectively. The diagonal components of the density matrices ($\rho_{ii}$, with $i=e, x$) represent the neutrino occupation numbers [in units of $1/\textrm{cm}^{3}$]; the number density of a given species, $n_{i}$, at a given location is obtained by $\int_{-1}^{1} \rho_{ii} d\cos\theta$. In Eqs.~\ref{vlasov1} and \ref{vlasov2}, the dependence of the density matrices on the polar angle $\theta$ (measured with respect to the radial direction, see Fig.~\ref{cartoon}) and radius $r$ is suppressed for the sake of brevity.

The velocity vector is represented by $\vec{v}$ while $\vec{\nabla}$ is the gradient operator. The inner product of the two is given by: 
\begin{eqnarray}
\vec{v}\cdot \vec{\nabla} = \cos\theta \frac{\partial}{\partial r} + \left(\frac{\sin^{2}\theta}{r}\right) \frac{\partial}{\partial \cos\theta}\ ,
\end{eqnarray}
where the dependence on the azimuthal angle $\phi$ is ignored due to the assumption of spherical symmetry.

The Hamiltonians for neutrinos and antineutrinos ($H = H_{\textrm{vac}} + H_{\textrm{mat}} + H_{\nu\nu} $ and $\bar{H} = -H_{\textrm{vac}} + H_{\textrm{mat}} + H_{\nu\nu}$, respectively) are made of the vacuum, matter, and the self-interaction terms:
\begin{eqnarray}
H_{\textrm{vac}} &=& \frac{\omega_{\rm{vac}}}{2}
\begin{pmatrix}
-\cos 2\theta_{\textrm{V}} & \sin 2\theta_{\textrm{V}} \\
\sin 2\theta_{\textrm{V}} & \cos 2\theta_{\textrm{V}}
\end{pmatrix}\ ,\\
H_{\textrm{mat}} &=&
\begin{pmatrix}
\lambda & 0 \\
0 & 0
\end{pmatrix}\ ,\\
H_{\nu\nu} &=& \sqrt{2} G_{\textrm{F}} \int d\cos\theta^{\prime} \left[\rho(\cos\theta^{\prime}) - \bar{\rho}(\cos\theta^{\prime})\right] \nonumber\\
& \times & (1-\cos\theta \cos\theta^{\prime})\ .
\end{eqnarray}
In the vacuum Hamiltonian, the vacuum frequency is $\omega_{\rm{vac}} = \Delta m^2/2\epsilon_\nu$, with $\Delta m^2 = 2.5 \times 10^{-3}$, and $\epsilon_\nu$ being the average neutrino energy calculated over all species, while the mixing angle is $\theta_{\textrm{V}} = 10^{-3}$. In the matter term, the strength of neutrino-matter interactions is given by $\lambda = \sqrt{2} G_{\textrm{F}} n_{e}$, with the electron number density being $n_{e}=Y_{e} \rho_{B}/m_N$ where $Y_e$ is the electron fraction, $\rho_{B}$ is the baryon mass density, and $m_N$ the nucleon mass (in the following, we restrict our analysis to the linear regime in which the matter term can be ignored because it does not affect the instability of the system). Note that, when we carry out the linear stability analysis in the following sections,  we neglect the matter term and effectively account for it though  a smaller value for the mixing angle (i.e., $\theta_{\textrm{V}} = 10^{-3}$). The strength of the self-interaction potential, generally denoted by $\mu = \sqrt{2} G_{\textrm{F}} n_{\nu_e}$, characterizes the $\nu$--$\nu$ term of the Hamiltonian.

The collision term, schematically represented by the functionals $\mathcal{C}$ and $\bar{\mathcal{C}}$, includes the absorption, emission, and direction changing processes defined as outlined in Appendix~\ref{collisions}: 
\begin{eqnarray}
\label{coll1}
\mathcal{C}[\rho, \bar\rho] &=& \mathcal{C}_{\textrm{emit}} - \mathcal{C}_{\textrm{absorb}} \odot \rho(\cos\theta)\nonumber \\ 
&+& \frac{\mathcal{C}_{\textrm{dir-ch}}}{2}\int d \cos\theta^\prime \left[-\rho(\cos\theta)+\rho(\cos\theta^{\prime})\right]\nonumber\\
&+& \cos\theta\ \mathcal{C}_{\textrm{ani}} \int d\cos\theta^{\prime} \cos\theta^{\prime} \rho(\cos\theta^{\prime})\ ,\\
\label{coll2}
\bar{\mathcal{C}}[\rho, \bar\rho] &=& \bar{\mathcal{C}}_{\textrm{emit}} - \bar{\mathcal{C}}_{\textrm{absorb}} \odot \bar{\rho}(\cos\theta)\nonumber \\ 
&+& \frac{\bar{\mathcal{C}}_{\textrm{dir-ch}}}{2}\int d\cos\theta^{\prime} \left[-\bar{\rho}(\cos\theta)+\bar{\rho}(\cos\theta^{\prime})\right]\nonumber\\
&+& \cos\theta\ \bar{\mathcal{C}}_{\textrm{ani}} \int d\cos\theta^{\prime} \cos\theta^{\prime} \bar{\rho}(\cos\theta^{\prime})\ .
\end{eqnarray}
Here, $\mathcal{C}_{\textrm{emit}}= \textrm{diag}(\mathcal{C}_{\textrm{emit}}^{ee}, \mathcal{C}_{\textrm{emit}}^{xx})$ and $\bar{\mathcal{C}}_{\textrm{emit}} = \textrm{diag}(\bar{\mathcal{C}}_{\textrm{emit}}^{ee}, \bar{\mathcal{C}}_{\textrm{emit}}^{xx})$ govern the emission rate for each neutrino flavor.
 $\mathcal{C}_{\textrm{absorb}}$ and $\bar{\mathcal{C}}_{\textrm{absorb}}$ are matrices and $\odot$ represents element-wise multiplication:
\begin{eqnarray}
\mathcal{C}_{\textrm{absorb}} \odot \rho(\cos\theta) &=& \mathcal{C}_{\textrm{absorb}}^{ee} \rho_{ee} + \mathcal{C}_{\textrm{absorb}}^{ex} \rho_{ex} \nonumber\\
&+& \mathcal{C}_{\textrm{absorb}}^{xe} \rho_{xe} + \mathcal{C}_{\textrm{absorb}}^{xx} \rho_{xx}\nonumber \\ 
\bar{\mathcal{C}}_{\textrm{absorb}} \odot \bar{\rho}(\cos\theta) &=& \bar{\mathcal{C}}_{\textrm{absorb}}^{ee} \bar{\rho}_{ee} + \bar{\mathcal{C}}_{\textrm{absorb}}^{ex} \bar{\rho}_{ex} \nonumber \\
&+& \bar{\mathcal{C}}_{\textrm{absorb}}^{xe} \bar{\rho}_{xe} + \bar{\mathcal{C}}_{\textrm{absorb}}^{xx} \bar{\rho}_{xx}\ ,\nonumber \\
\end{eqnarray}
which is linearly proportional to the number of neutrinos at a given location representing the rate at which neutrinos are absorbed by the surrounding matter~\cite{Richers:2019grc}. The off-diagonal components of $\mathcal{C}_{\textrm{absorb}}$ and $\bar{\mathcal{C}}_{\textrm{absorb}}$ are the average of the diagonal elements of the diagonal terms. 
The terms $\mathcal{C}_{\textrm{dir-ch}}$ and $\bar{\mathcal{C}}_{\textrm{dir-ch}}$ 
describe the direction changing neutral current scatterings, which do not change the number of neutrinos and are flavor independent. 
The $\mathcal{C}_{\rm{ani}}$ and $\bar{\mathcal{C}}_{\rm{ani}}$ terms depend on the proton and neutron fractions. 
Moreover, we take into account the anisotropy in the direction changing scatter term. 

In Eqs.~\ref{coll1} and \ref{coll2}, the emission and absorption terms seem to be independent of each other, however, that is not the case and they are related by the Kirchoff's law as explained in Appendix~\ref{collisions}. Hence, given either the absorption or the emission rate, the other can be calculated (for this reason, in Appendix~\ref{collisions}, we only provide the absorption rates for all the processes).

\section{Steady state neutrino angular distributions}
\label{sec:steady}
\begin{figure*}
\includegraphics[width=0.99\textwidth]{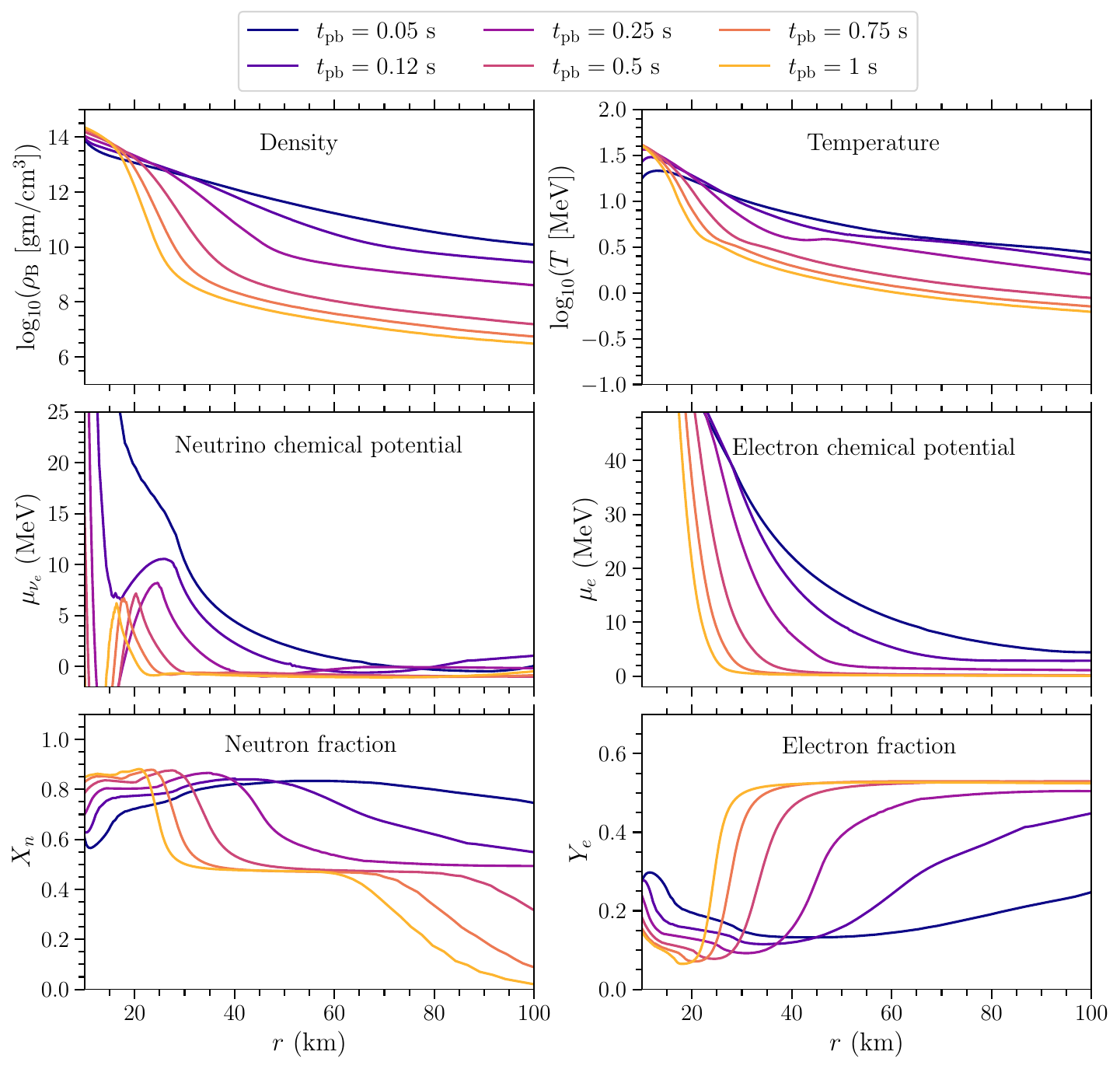}
\caption{Radial profiles of the characteristic  quantities extracted from our benchmark hydrodynamical SN simulation (see main text for details)  at the post-bounce times $t_{\rm{pb}} = 0.05$, $0.12$, $0.25$, $0.5$, $0.75$, and $1$~s and then adopted to compute the neutrino angular distributions. The different panels represent the baryon number density, the electron temperature, the neutrino and electron chemical potentials, and the neutron and electron fractions, from top left to bottom right, respectively. 
}
\label{thermo}
\end{figure*}

We adopt the outputs of a one-dimensional hydrodynamical simulation of a SN with mass $18.6\ M_\odot$, SFHo nuclear equation of state without muons, and gravitational mass $1.4\ M_\odot$~\cite{SNarchive}. We consider static hydrodynamical backgrounds and thermodynamical quantities, as well as the local neutrino number density and first energy moment, for selected time snapshots extracted at post-bounce times $t_{\rm{pb}} = 0.05$, $0.12$, $0.25$, $0.5$, $0.75$, and $1$~s, which are representative of the accretion as well as the early Kelvin-Helmoltz cooling phases. Then, we use these quantities as inputs to calculate the collision terms (Eqs.~\ref{coll1} and \ref{coll2}). The relevant properties of our benchmark hydrodynamical SN simulations adopted to compute the collision terms are illustrated in Fig.~\ref{thermo} (see also Appendix~\ref{collisions}).

The neutrino angular distributions are not provided by the hydrodynamic simulation, hence, we compute them  solving Eqs.~\ref{vlasov1} and \ref{vlasov2} and imposing $H=\bar{H}=0$. Since we are only interested in the angular distributions in the proximity of the decoupling region, we solve Eqs.~\ref{vlasov1} and \ref{vlasov2} in a radial shell (see Fig.~\ref{cartoon}) whose width $[r_{\rm{min}}, r_{\rm{max}}]$ differs for each post-bounce time as summarized in Table~\ref{table:rminrmax} and has been chosen such that the minimum radius $r_{\rm min}$ has a number density of neutrinos that coincides with the energy integrated Fermi-Dirac distribution extracted from the hydrodynamical simulation (i.e., at large optical depth, such that neutrinos are fully coupled with matter), while neutrinos are completely decoupled at the maximum radius $r_{\rm max}$ (i.e., the backward component of the neutrino and antineutrino number densities is negligible: $\rho_{ii}(\cos\theta) \approx 0$ for $\cos\theta<0$). %This is to ensure that we can reliably impose the boundary condition $\rho = 0$ for $\cos\theta<0$. 
We solve the neutrino kinetic equations starting with no neutrinos and antineutrinos in the shell. As the system evolves in time, the simulation shell is populated with neutrinos and antineutrinos with the boundary conditions being guided by the hydrodynamical simulation as explained above.

\begin{table}
\label{table:rminrmax}
\caption{Values of $r_{\textrm{min}}$ and $r_{\textrm{min}}$ used for each post-bounce time snapshot to take into account the varying location and width of the decoupling region (see also Fig.~\ref{cartoon}).}
\begin{tabular}{|l|l|l|}
\hline
$t_{\rm{pb}}$ (s) & $r_{\textrm{min}}$ (km) & $r_{\textrm{min}}$ (km) \\
\hline
\hline
$0.05$ & 25 & 150 \\
\hline
$0.12$ & 25 & 100 \\
\hline
$0.25$ & 22 & 57 \\
\hline
$0.5$ & 20 & 35 \\
\hline
$0.75$ & 17 & 32 \\
\hline
$1$ & 16 & 31\\
\hline
\end{tabular}
\end{table}

For each time snapshot, we use a grid with $75$ radial bins and $75$ angular bins uniformly spread over $\cos\theta$ and
solve the neutrino kinetic equations for the time interval required to achieve a classical steady state configuration. 

\begin{figure*}
\includegraphics[width=0.99\textwidth]{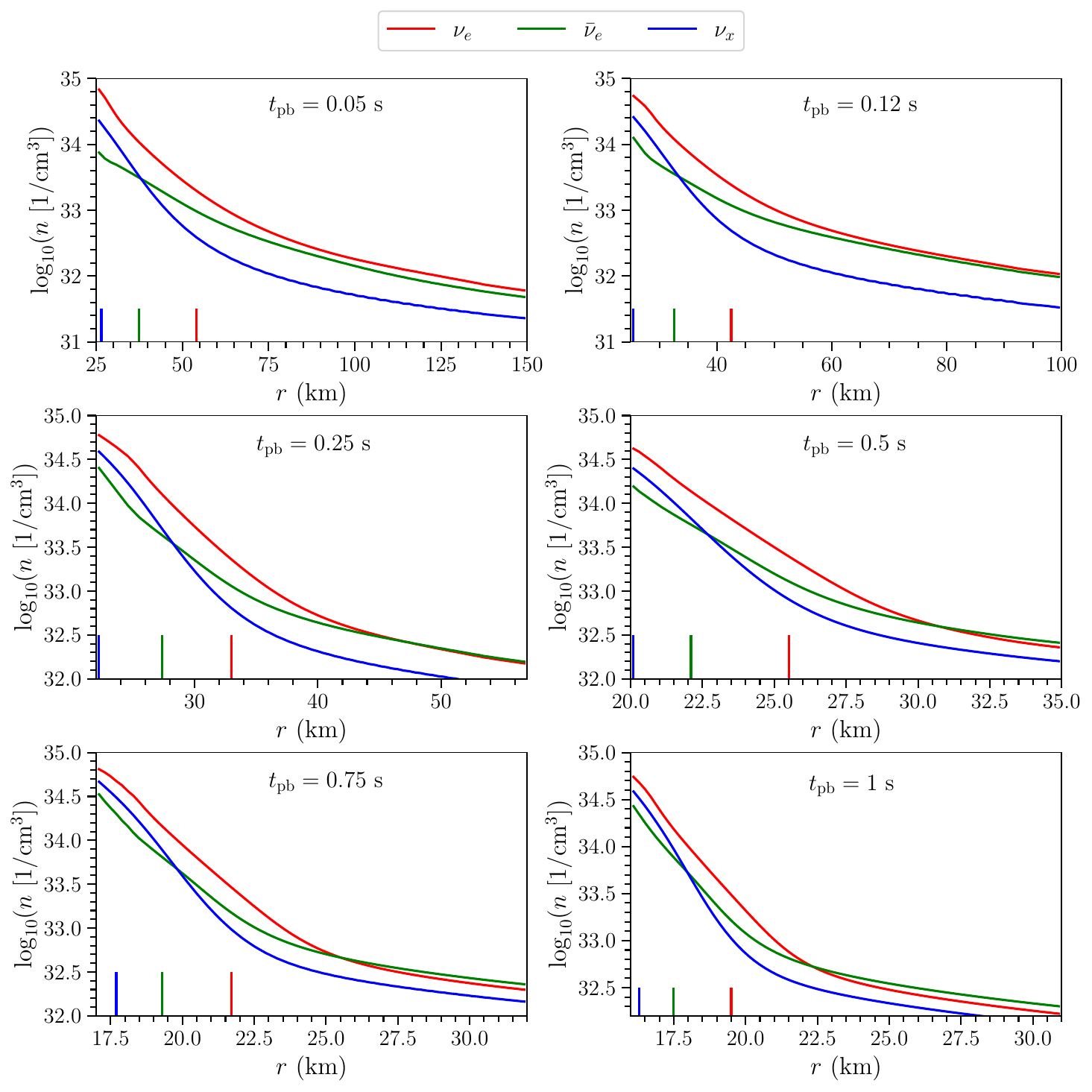}
\caption{Radial evolution of the number densities of $\nu_{e}$ (in red), $\bar{\nu}_{e}$ (in green), and $\nu_{x}$ (in blue) for  selected post-bounce time snapshots and computed solving the neutrino kinetic equations. We mark the radius of neutrino decoupling through a vertical line; this is the radius at which the number density deviates from a Fermi-Dirac distribution (see Appendix~\ref{sec:comparison}). One can see that the decoupling surface shrinks as a function of time. Note that the radial range is not the same for all time snapshots.}
\label{numden}
\end{figure*}

Figure~\ref{numden} shows the radial profiles of the angle-integrated neutrino number densities computed for all three flavors through the procedure described above. While involving some approximations, our method allows us to reproduce a local neutrino density in agreement with the output of our benchmark hydrodynamical simulation, as shown in Appendix~\ref{sec:comparison}. We note that, due to the single energy approximation used in this paper for simplicity, we slightly overestimate the number density of $\bar{\nu}_{e}$ with respect to the one of $\nu_{e}$ at large radii when comparing with our benchmark hydrodynamical simulation (see also Fig.~\ref{numden1}). As shown later, the fast flavor instability is triggered by the presence of ELN crossings that would be still present, despite the single-energy approximation; hence,  the relative importance of fast flavor instabilities with respect to the collisional ones is not affected by the single-energy treatment.
 We stress that the goal of the procedure illustrated in this section is not to reproduce the neutrino properties obtained through the hydrodynamical simulation, rather to obtain neutrino angular distributions that are not arbitrary but inspired by conditions found in the SN interiors.

We define the decoupling radius as the one at which the neutrino number density differs from the Fermi-Dirac one up to $1\%$ (see Appendix~\ref{sec:comparison}) and mark such radius in Fig.~\ref{numden} to guide the eye. One can see from Fig.~\ref{numden} that the decoupling radius becomes smaller as the post-bounce time increases. 
Because of the different nature of their interactions, while the electron flavors decouple over a broader radial range, the non-electron flavor decouples almost instantaneously~\cite{Tamborra:2017ubu}. 

\begin{figure}
\includegraphics[width=0.35\textwidth]{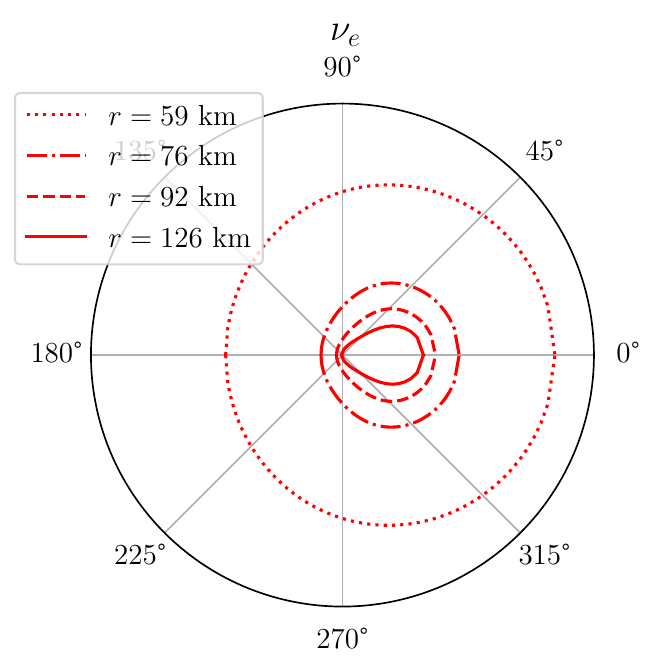}\\
\includegraphics[width=0.35\textwidth]{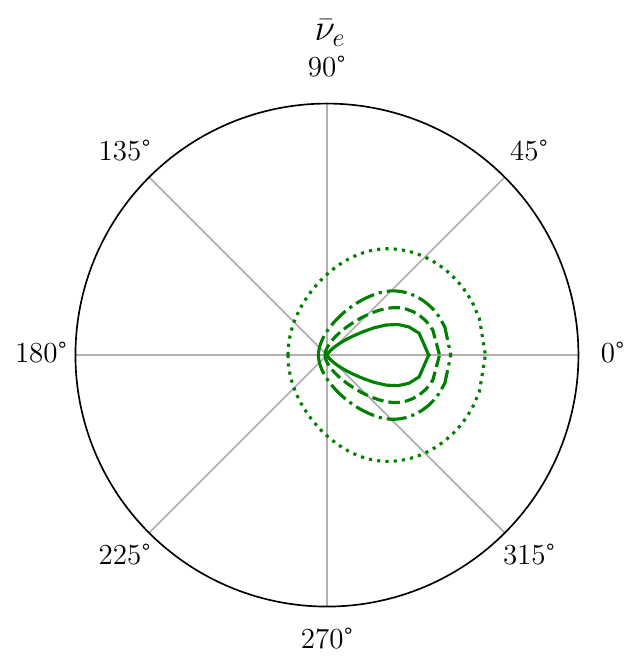}\\
\includegraphics[width=0.35\textwidth]{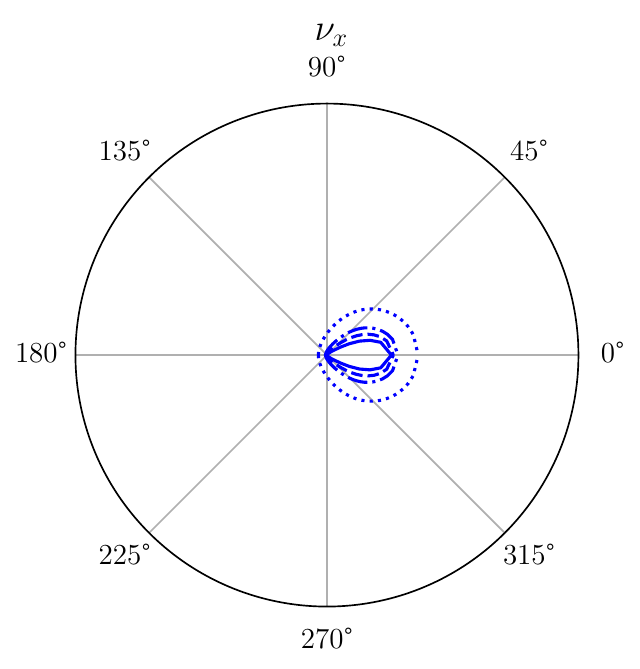}
\caption{Polar diagram of the radial variation of the spectral intensity of $\nu_e$, $\bar\nu_e$, and $\nu_x$, from top to bottom respectively,  computed solving the neutrino kinetic equations. 
The selected radii and the time snapshot $t_{\rm pb}=0.05$~s have been chosen for illustrative purposes. One can see that the neutrino distributions become progressively forward peaked as $r$ increases, with an angular spread that is largest for $\nu_e$, followed by $\bar\nu_e$, and then $\nu_x$. }
\label{fig:polar}
\end{figure}

Figure~\ref{fig:polar} displays an example of the radial evolution of the neutrino angular distributions computed through the procedure illustrated above for $t_{\rm pb} = 0.05$~s. As also found in Ref.~\cite{Tamborra:2017ubu}, the angular distributions become progressively forward peaked as a function of the radial distance and have an angular spread that is the largest for $\nu_e$, followed by $\bar\nu_e$ and $\nu_x$. 

\begin{figure}
\includegraphics[width=0.45\textwidth]{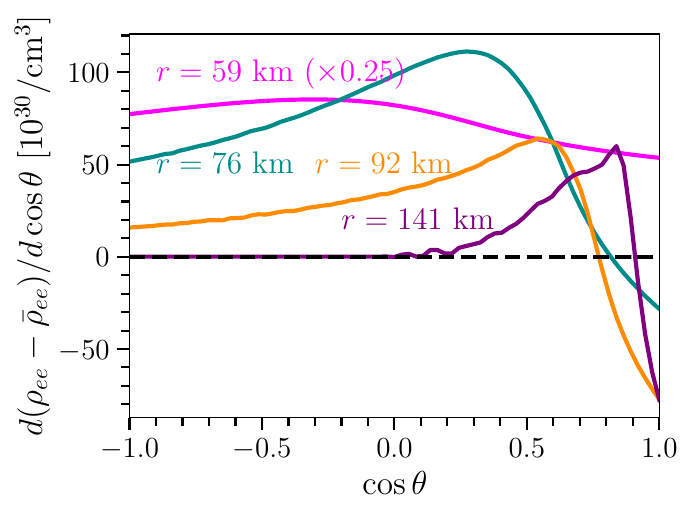}\\
\caption{Angular distribution of the ELN for $t_{\rm pb}=0.05$~s extracted at different radii as indicated. We find ELN crossings for all considered post-bounce time snapshots. }
\label{fig:ELN}
\end{figure}

Figure~\ref{fig:ELN} focuses on the radial evolution of the ELN angular distribution for a representative post-bounce time snapshot ($t_{\rm pb}=0.05$~s), 
interestingly, we find ELN crossings for most radii. This finding holds for all post-bounce times considered in this work for our benchmark SN model and is in contrast with the results presented in Ref.~\cite{Tamborra:2017ubu}; the latter concluded, through a systematic analysis of different one-dimensional SN models, that ELN crossings are not present in one-dimensional hydrodynamical simulations, albeit ELN crossings were found in multi-dimensional SN simulations~\cite{Morinaga:2019wsv,Abbar:2019zoq,Abbar:2018shq,DelfanAzari:2019tez,Nagakura:2019sig,Abbar:2020qpi}.
The results presented in Fig.~\ref{fig:ELN} seem to fulfill the criterion of Ref.~\cite{Shalgar:2019kzy} that linked the occurrence of ELN crossings to comparable number densities of neutrinos and antineutrino (see also Refs.~\cite{Lin:2022dek,Kato:2023dcw}). Yet, the reason why ELN crossings are found in this work, but not in Ref.~\cite{Tamborra:2017ubu} requires further investigation and will be the focus of an upcoming dedicated paper~\cite{in_prep}---we stress that the purpose of this work is to rely on the angular distributions derived from first principles to investigate the interplay among fast, slow, and collisional instabilities. The different findings could be connected to the fact that the proto-neutron star cooling model of our benchmark SN simulation includes effects from proto-neutron star convection (treated by a mixing-length approximation~\cite{Mirizzi:2015eza,2004cgps.book.....W}) in contrast to the SN models investigated in Ref.~\cite{Tamborra:2017ubu}. Other reasons might be the different nuclear equation of state (SFHo for our benchmark SN model vs.~Lattimer \& Swesty with compressibility modulus $K = 220$~MeV in Ref.~\cite{Tamborra:2017ubu}) or secondary subtle differences in our neutrino transport results compared to those obtained with the VERTEX transport used in Ref.~\cite{Tamborra:2017ubu}. 

\section{Linear stability analysis for the homogeneous mode}
\label{sec:lsa}

The radial range where significant flavor evolution might develop can be determined through the linear stability analysis~\cite{Banerjee:2011fj,Airen:2018nvp,Izaguirre:2016gsx,Morinaga:2018aug}. In this section, we linearize Eqs.~\ref{vlasov1} and \ref{vlasov2} assuming homogeneity (i.e.,~ignoring the advective term in the neutrino kinetic equations) and perform the linear stability analysis with and without the collision term to investigate the impact of collisional instabilities. Since we consider monoenergetic neutrinos for simplicity, we here assume that the neutrino energy $\epsilon_\nu$ entering $\omega_{\rm{vac}}$ changes as a function of the radius and it coincides with the average of the first energy moments extracted from the hydrodynamical SN simulation over all four flavors (see bottom right panel of Fig.~\ref{fig:Eave}).

The linearized version of Eqs.~\ref{vlasov1} and \ref{vlasov2} for the off-diagonal terms of the density matrices at a given $r$ and $\cos\theta$ is
\begin{eqnarray}
\label{lin1}
i\frac{\partial \rho_{ex}}{\partial t} &=& \left[ (H_{ee}-H_{xx})\rho_{ex} -(\rho_{ee}-\rho_{xx})H_{ex}\right] \nonumber\\
&-& \frac{i}{2}(\mathcal{C}^{e}_{\textrm{absorb}} + \mathcal{C}^{x}_{\textrm{absorb}})\rho_{ex}\nonumber\\
&-& \frac{i}{4}(\mathcal{C}^{e}_{\textrm{dirch}} + \mathcal{C}^{x}_{\textrm{dir-ch}}) \nonumber \\
&\times& \int [\rho_{ex}(\cos\theta)-\rho_{ex}(\cos\theta^{\prime})] d\cos\theta^{\prime}\ ,\\
\label{lin2}
i\frac{\partial \bar{\rho}_{ex}}{\partial t} &=& \left[ (\bar{H}_{ee}-\bar{H}_{xx})\bar{\rho}_{ex} -(\bar{\rho}_{ee}-\bar{\rho}_{xx})\bar{H}_{ex} \right]\nonumber\\
&-& \frac{i}{2}(\bar{\mathcal{C}}^{e}_{\textrm{absorb}} + \bar{\mathcal{C}}^{x}_{\textrm{absorb}})\bar{\rho}_{ex}\nonumber\\
&-& \frac{i}{4}(\bar{\mathcal{C}}^{e}_{\textrm{dirch}} + \bar{\mathcal{C}}^{x}_{\textrm{dir-ch}}) \nonumber \\
&\times& \int [\bar{\rho}_{ex}(\cos\theta)-\bar{\rho}_{ex}(\cos\theta^{\prime})] d\cos\theta^{\prime}\ ,
\end{eqnarray}

It should be noted that, due to the direction changing contribution in the collision term, the linearized equations for $\rho_{ex}(r,\cos\theta)$ and $\bar{\rho}_{ex}(r,\cos\theta)$ depend on the density matrix at all other angles, due to the integral in the last term of Eqs.~\ref{lin1} and \ref{lin2}.

Equations \ref{lin1} and \ref{lin2} form an eigensystem for which we calculate the eigenvalues following the procedure described in Ref.~\cite{Banerjee:2011fj}. In the absence of the collision term, the eigenvalues of the system are complex conjugate pairs; however, due to the presence of the collision term this is no longer true for Eqs.~\ref{lin1} and \ref{lin2}. The eigenmode that has the eigenvalue with the largest positive imaginary term, which we denote by $\kappa$, dominates and is responsible for an instability for which we can compute its growth rate. 

The growth rate of fast, slow, and collisional flavor instabilities is linked to three time scales of our system, $\mu^{-1}$, $\omega_{\textrm{vac}}^{-1}$, and the time between two successive collisions of neutrinos--see also Refs.~\cite{Shalgar:2022lvv,Shalgar:2022rjj}. 
The time scales associated with the growth rate is determined by the fastest time scale that contributes to the flavor instability.

\begin{figure*}
\includegraphics[width=0.49\textwidth]{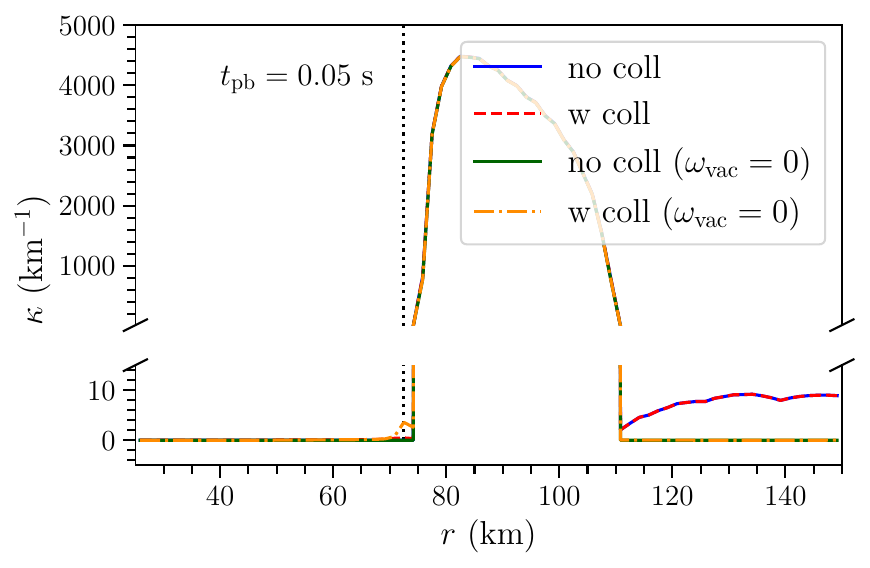}
\includegraphics[width=0.49\textwidth]{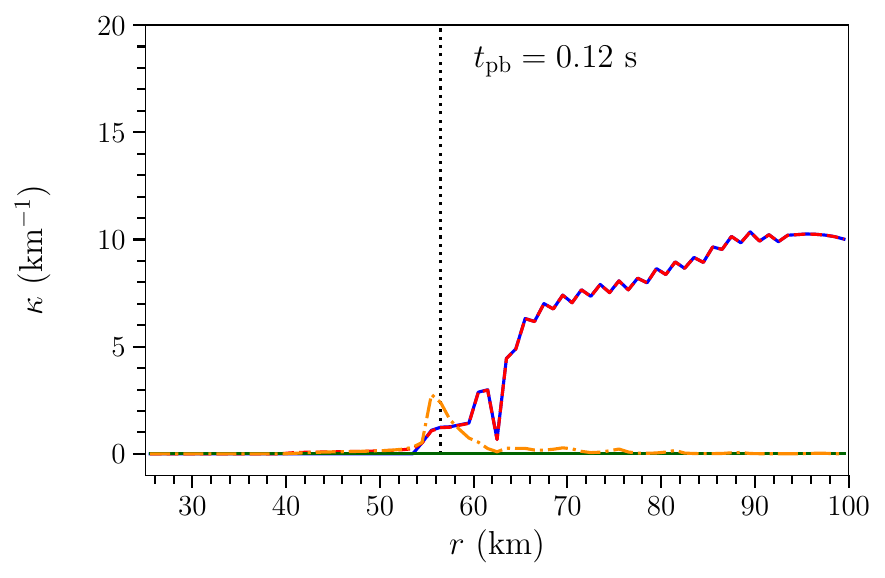}
\includegraphics[width=0.49\textwidth]{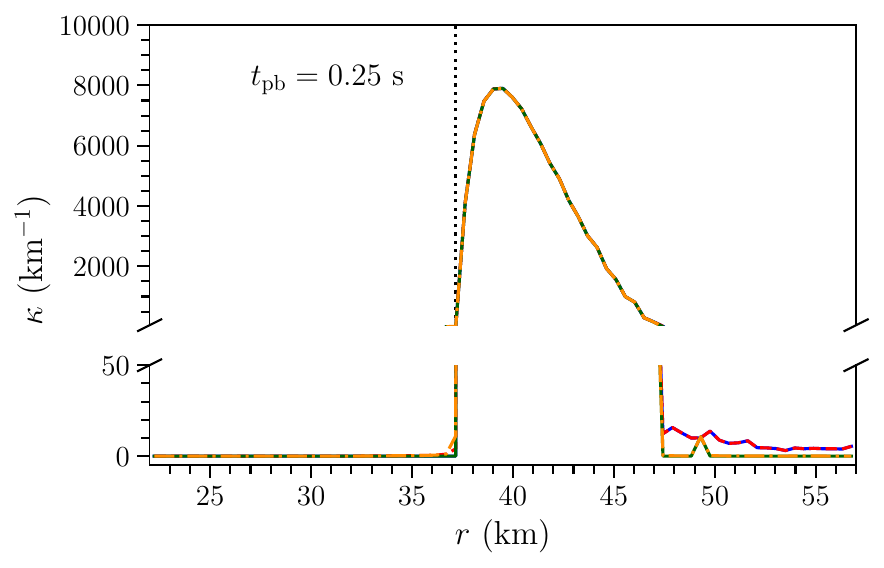}
\includegraphics[width=0.49\textwidth]{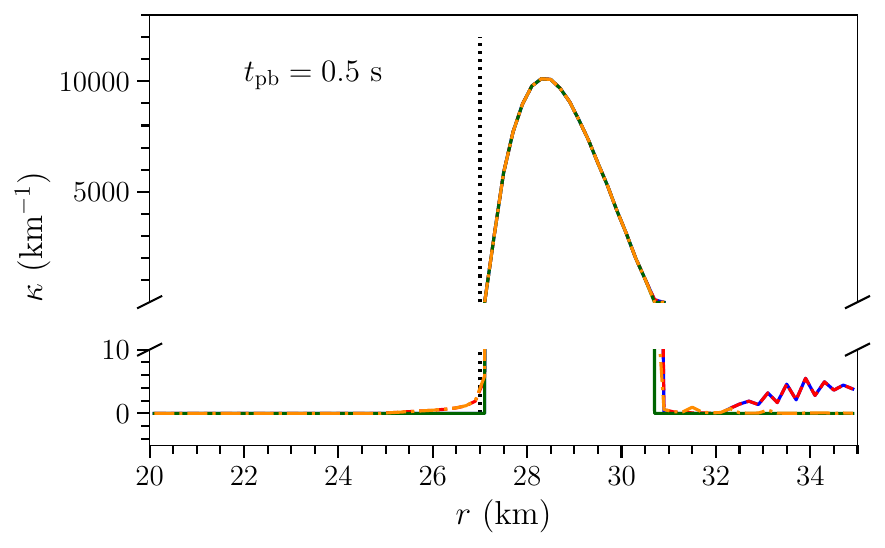}
\includegraphics[width=0.49\textwidth]{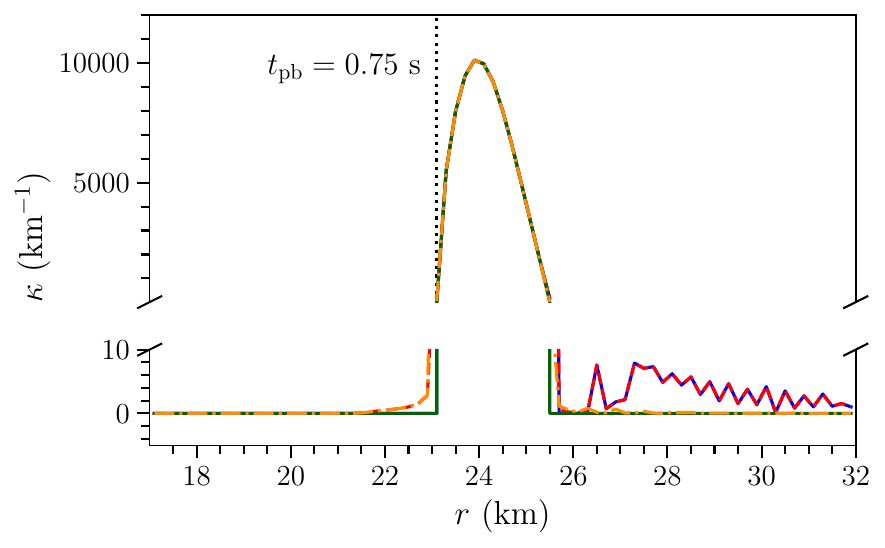}
\includegraphics[width=0.49\textwidth]{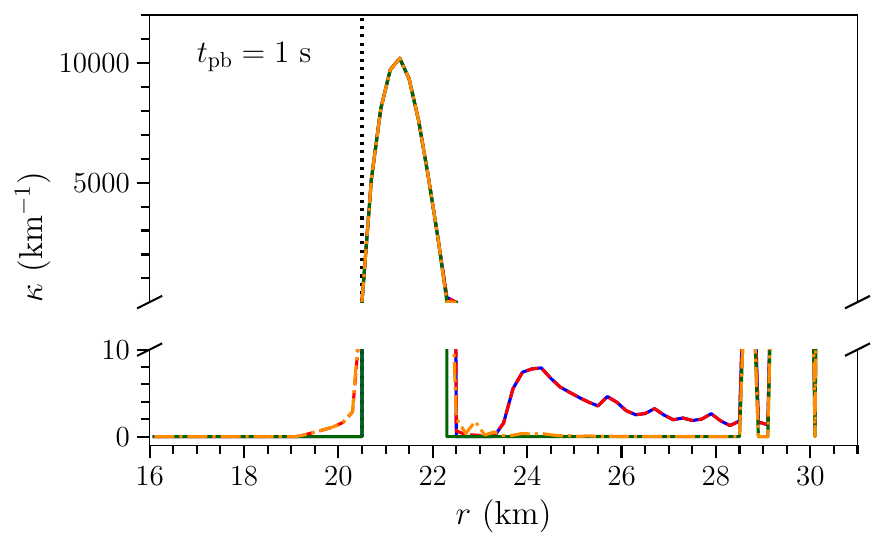}
\caption{Radial profile of the growth rate of the flavor instability, obtained under the assumption of homogeneity, at various time snapshots (from $t_{\rm pb}=0.05$~s in the top left panel to $t_{\rm pb}=1$~s in the bottom right panel). Each panel displays results obtained for four different scenarios. The blue line uses the vacuum and the self-interaction Hamiltonian (slow instability), while the green line only considers the self-interaction Hamiltonian (fast instability); for both blue and green lines, collisional damping is ignored. The same calculation is repeated to obtain the dashed red and dot-dashed orange lines, but with collisional damping. It should be noted that, except for $t_{\rm pb}=0.12$~s, the four lines are almost on top of each other and difficult to distinguish. The location of the onset of ELN crossings is marked through the black vertical line to guide the eye.}
\label{growth}
\end{figure*}
Figure~\ref{growth} shows the growth rates for the fast instability (obtained by calculating the eigenvalues of Eqs.~\ref{lin1} and \ref{lin2} when $\omega_{\textrm{vac}}=0$, green curve), along with the growth rates obtained when the vacuum and the collision terms are included. 
We find regions of flavor instability for all time snapshots considered in this work. Interestingly, in all cases except for $t_{\textrm{pb}}=0.12$~s, the largest growth rate is triggered by the ELN crossing (green curve),
with regions (especially at larger radii) where instabilities due to $\omega_{\textrm{vac}} \neq 0$ dominate (blue and dashed red curves), but with a growth rate which is about three orders of magnitude smaller than the first peak.
 The behavior of the growth rate for $t_{\rm pb}=0.12$~s represents a special case (top right panel of Fig.~\ref{growth}); in fact, the fast instability is absent (green line) in spite of the presence of an ELN crossing. It is the interplay between the vacuum frequency with the ELN crossing that makes the system unstable and leads to the growth rate of the blue curve.  This is confirmed by the smaller (slow) growth rate that we observe in this case with respect to all other snapshots. The fact that the blue and red curves are exactly on top of each other implies a subleading role of the collisional instability at $r \gtrsim 56$~km. The effect of the collisional instability is visible for $\omega_{\textrm{vac}}=0$ (dot-dashed orange curve).

\begin{figure}
\includegraphics[width=0.49\textwidth]{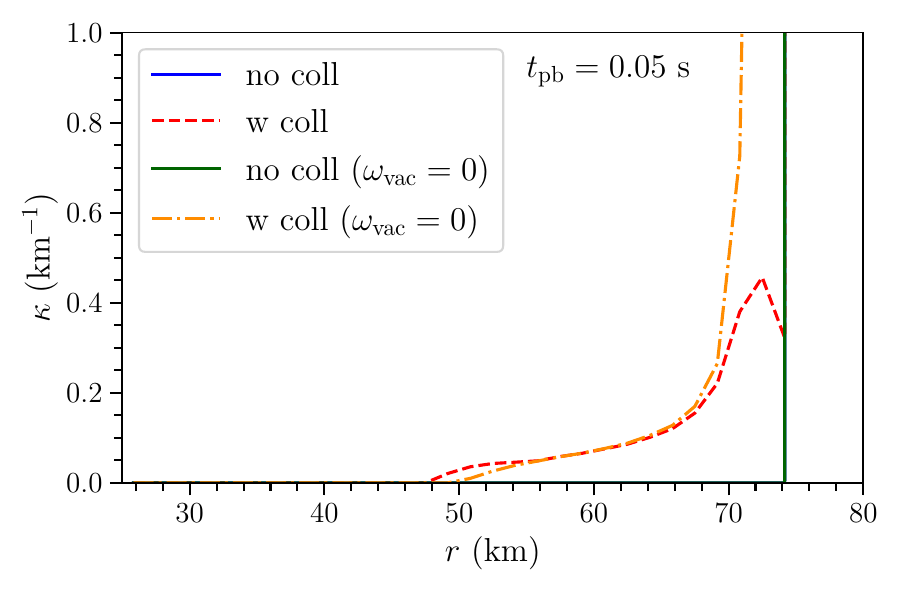}
\includegraphics[width=0.49\textwidth]{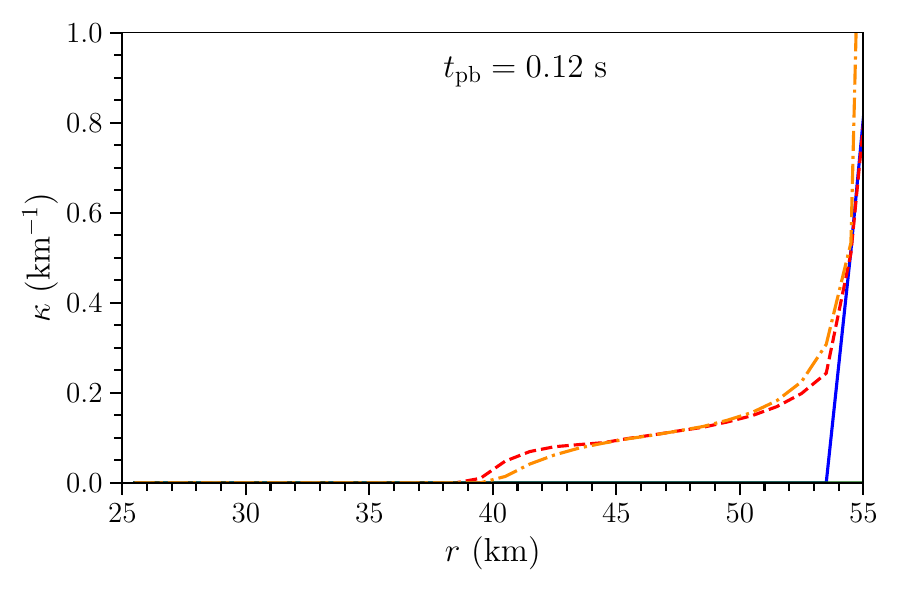}
\includegraphics[width=0.49\textwidth]{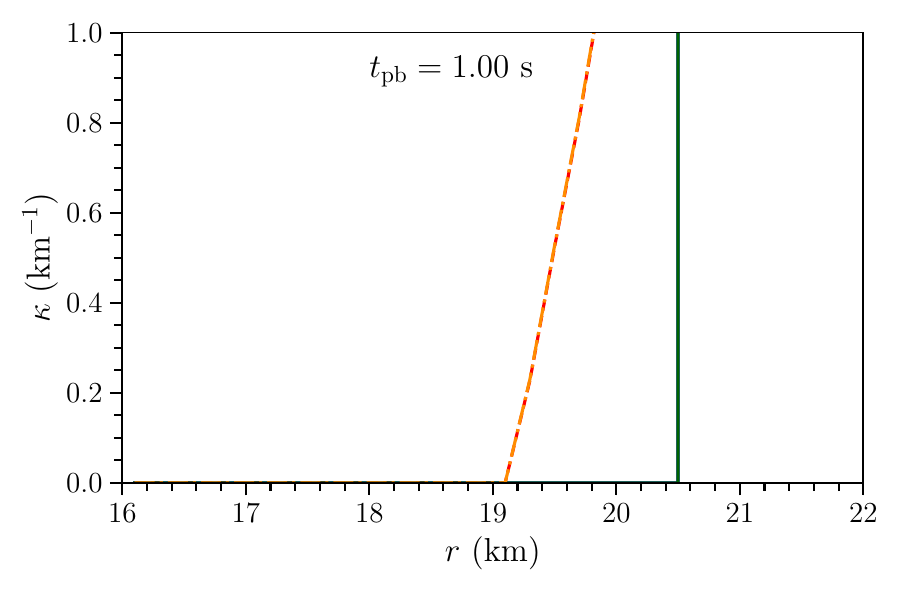}
\caption{Zoomed in version of Fig.~\ref{growth} for $t_{\rm pb}=0.05$, $0.12$, and $1$~s to highlight  the radial region where the collisional instability starts to be present. In the top and the bottom panels, the blue line is not visible as it is behind the green line.}
\label{growthzoom}
\end{figure}

We conclude that the growth rate does not seem to be dominated by collisional effects (dot-dashed orange curve) across the investigated radial range and for all times, except for a few minor exceptions. 
In very small regions just prior to the growth of $\kappa$ of the fast flavor instability, the collisional instability is found, albeit with a very small growth rate compared to the fast flavor instability, the collisional instability having the effect of pushing the occurrence of the flavor instability to a smaller radius by a few kilometers. Hence, collisions seem to modify the effective vacuum frequency, in agreement with the findings of Ref.~\cite{Xiong:2022zqz}.
 In order to better inspect the growth of the collisional instability,  Fig.~\ref{growthzoom} shows a zoomed in version of Fig.~\ref{growth} for $t_{\rm{pb}}=0.05$, $0.12$, and $1$~s. We can clearly see a small collisional growth rate that appears just prior to the fast or slow instability; however, the impact of this small instability is likely to be negligible. The reason for this is that the flavor instability develops before neutrinos are completely decoupled from matter, and hence forward peaked. The flavor evolution at larger radii can thus be transported to smaller radii because of neutrino advection; hence, the fast or slow flavor instability might still have a larger impact than collisional instability on the flavor evolution at smaller radii.
These findings suggest that, if the growth rate of slow and fast instabilities should always be larger than the collisional one, then the only scenario where the collisional instability might be important should be when both fast and slow flavor instability are absent.

It should be emphasized that the presence of a flavor instability in a limited radial range does not imply that significant flavor conversion develops in or is restricted to that region; in fact, the advective term can cause the flavor content at other radii to be modified because of flavor conversion, even if the flavor instability is not initially present in a certain spatial range~\cite{Shalgar:2019qwg,Wu:2021uvt,Nagakura:2022xwe,Shalgar:2022lvv,Shalgar:2022rjj}

\section{Impact of inhomogeneity on the development of the flavor instability}
\label{sec:lsa_inh}

In this section, we address the effect of inhomogeneities on the results presented in Sec.~\ref{sec:lsa}. 
To this purpose, we generalize Eqs.~\ref{lin1} and \ref{lin2} to include the spatial dependence as follows:
\begin{eqnarray}
\label{lin1sp}
i\frac{\partial \rho_{ex}}{\partial t} &=& -i\vec{v}\cdot\vec{\nabla}\rho_{ex} + \left[ (H_{ee}-H_{xx})\rho_{ex} -(\rho_{ee}-\rho_{xx})H_{ex}\right] \nonumber\\
&-& \frac{i}{2}(\mathcal{C}^{e}_{\textrm{absorb}} + \mathcal{C}^{x}_{\textrm{absorb}})\rho_{ex}\nonumber\\ 
&-& \frac{i}{4}(\mathcal{C}^{e}_{\textrm{dirch}} + \mathcal{C}^{x}_{\textrm{dir-ch}}) \nonumber \\
&\times& \int (\rho_{ex}(\cos\theta)-\rho_{ex}(\cos\theta^{\prime})) d\cos\theta^{\prime}\ ,
\end{eqnarray} 
\begin{eqnarray}
\label{lin2sp}
i\frac{\partial \bar{\rho}_{ex}}{\partial t} &=& -i\vec{v}\cdot\vec{\nabla}\bar{\rho}_{ex} + \left[ (\bar{H}_{ee}-\bar{H}_{xx})\bar{\rho}_{ex} -(\bar{\rho}_{ee}-\bar{\rho}_{xx})\bar{H}_{ex} \right]\nonumber\\
&-& \frac{i}{2}(\bar{\mathcal{C}}^{e}_{\textrm{absorb}} + \bar{\mathcal{C}}^{x}_{\textrm{absorb}})\bar{\rho}_{ex}\nonumber\\
&-& \frac{i}{4}(\bar{\mathcal{C}}^{e}_{\textrm{dirch}} + \bar{\mathcal{C}}^{x}_{\textrm{dir-ch}}) \nonumber \\
&\times& \int (\bar{\rho}_{ex}(\cos\theta)-\bar{\rho}_{ex}(\cos\theta^{\prime})) d\cos\theta^{\prime}\ .
\end{eqnarray} 

It can be seen that the growth rate of $\rho_{ex}$ and $\bar{\rho}_{ex}$ as a function of time at a given spatial location can be modified by the spatial derivative only if $\vec{\nabla}\rho_{ex}$ and $\vec{\nabla}\bar{\rho}_{ex}$ are significant in comparison to the other terms entering the kinetic equations.

If periodic boundaries are assumed, in order to perform the linear stability analysis for the inhomogeneous modes, the ansatz that $\rho_{ex}$ and $\bar{\rho}_{ex}$ can be represented as a combination of Fourier modes, i.e.~$\rho_{ex}(\vec{x}) = \sum_{\vec{k}} e^{i\vec{k}\cdot\vec{x}} \tilde{\rho}_{ex}^{\vec{k}}$ and 
$\bar{\rho}_{ex}(\vec{x}) = \sum_{\vec{k}} e^{i\vec{k}\cdot\vec{x}} \bar{\tilde{\rho}}_{ex}^{\vec{k}}$, has been usually considered in the literature~\cite{Duan:2014gfa, Abbar:2015mca}.
Here, $\tilde{\rho}_{ex}^{\vec{k}}$ and $\bar{\tilde{\rho}}_{ex}^{\vec{k}}$ are the Fourier components of the off-diagonal components of the density matrices. 
Plugging the expressions for $\rho_{ex}(\vec{x})$ and $\bar{\rho}_{ex}(\vec{x})$ in Eqs.~\ref{lin1sp} and \ref{lin2sp}, one can see that the linearized equations decouple for each Fourier mode, if and only if the Hamiltonian is independent of the spatial location, i.e.~$\rho_{ee}$ and $\bar{\rho}_{ee}$ are not functions of $\vec{x}$. If the Hamiltonian has non-trivial spatial dependence, as in our case or the one investigated in Ref.~\cite{Xiong:2022vsy}, this standard approach cannot be applied to investigate the growth of the inhomogeneous modes~\footnote{Note that the method adopted in the original papers proposing the linear stability analysis for the inhomogeneous modes was justified since their goal was to explore the spontaneous breaking of spherical symmetry, without inhomogeneity in radial direction~\cite{Duan:2014gfa, Abbar:2015mca}}.

To perform the linear stability analysis in the presence of inhomogeneous modes, we should seek the eigenvalues of Eqs.~\ref{lin1sp} and \ref{lin2sp} for all spatial points simultaneously. Since this task is challenging, we look for the eigenvalues through an iterative procedure. 
We begin with the ansatz of collective evolution:
\begin{eqnarray}
\label{rhoform}
\rho_{ex}(\vec{x}) & \sim & Q_{\theta}(r) e^{-i\Omega(r)t}\ , \\
\label{rhobarform}
\bar{\rho}_{ex}(\vec{x}) & \sim & \bar{Q}_{\theta}(r) e^{-i\Omega(r)t}\ .
\end{eqnarray}
Our goal is to find the functions $\Omega(r)$, $Q_{\theta}(r)$, and $\bar{Q}_{\theta}(r)$ that satisfy Eqs.~\ref{lin1sp} and \ref{lin2sp}. We find the zeroth order approximation for these functions, ignoring the advective term; note that $\Omega^{0}(r)$, $Q_{\theta}^{0}(r)$, and $\bar{Q}_{\theta}^{0}(r)$ are independent of the spatial derivative in the zeroth order approximation. For each given location, the zeroth order estimate of $\Omega(r)$ denoted by $\Omega^{0}(r)$ is obtained requiring that
\begin{eqnarray}
\label{thedet}
\begin{vmatrix}
I[1]-1 & -I[\cos\theta] \\
I[\cos\theta] & -I[\cos^{2}\theta]-1
\end{vmatrix} = 0\ , 
\end{eqnarray}
where
\begin{eqnarray}
\label{Idef}
I[f] = \int d \cos\theta \frac{\rho_{ee}-\bar{\rho}_{ee}}{\Omega^{0}-H_{ee}}\ . 
\end{eqnarray}
Once the eigenvalue is known, the zeroth order estimate of the eigenvectors $Q^{0}_{\theta}(r)$ and $\bar{Q}^{0}_{\theta}(r)$ can be computed.
With the knowledge of the approximate dependence of $Q_{\theta}(r)$ and $\bar{Q}_{\theta}(r)$ on $\cos\theta$ and $r$, it is possible to estimate the magnitude of the advective term in Eqs.~\ref{lin1sp} and \ref{lin2sp}. This helps us to obtain an iterative estimate of the growth rate solving Eq.~\ref{thedet}, but with the denominator that is modified to include the advective term from the previous iteration. 

The advective term applied to Eqs.~\ref{rhoform} and \ref{rhobarform} gives
\begin{eqnarray}
\label{advecterm}
\vec{v} \cdot \nabla \rho_{ex} = \left(\frac{1}{Q_{\theta}(r)}\vec{v} \cdot \nabla Q_{\theta}(r) - i\frac{d\Omega(r)}{dr} \right) \rho_{ex}\ ;
\end{eqnarray}
and an analogous expression holds for antineutrinos. The first term in the parenthesis is generally small compared to the second term, but depends on 
$\cos\theta$. 
The effect of the advective term is to replace $\Omega^{0}$ in Eq.~\ref{Idef} with $(\Omega^{0}-i {d\Omega}/{dr})$. The real part of ${d\Omega}/{dr}$ is equivalent to adding a matter term that does not affect the growth rate of the instability. The imaginary part of ${d\Omega}/{dr}$ shifts the imaginary part of the eigenvalue. This correction is not significant if the derivative is small compared to $\textrm{Im}(\Omega^{0})$. For all the cases, we find that the results of this iterative approach lead to results that are  comparable to the ones shown in Fig.~\ref{growth} and hence are not shown.

\begin{figure}
\includegraphics[width=0.49\textwidth]{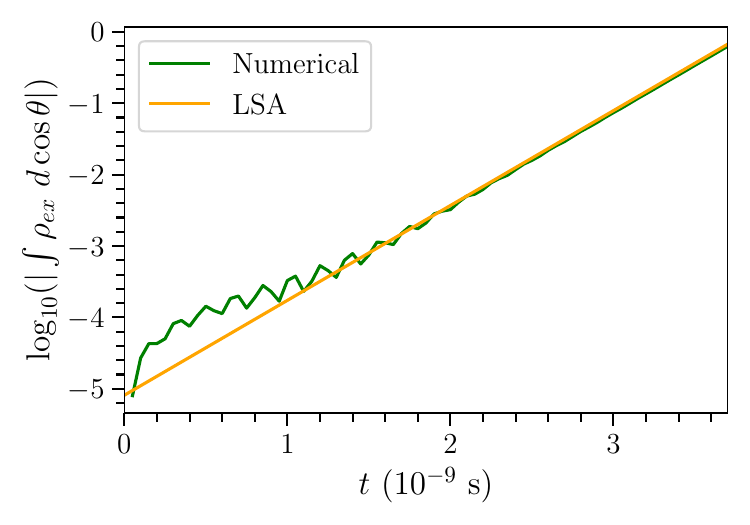}
\caption{Growth of $\left|\int \rho_{ex}~d\cos\theta\right|$  as a function of time at $r=21.3$ km  for $t_{\textrm{pb}}=1$~s. The growth rate is given by the slope of $\left|\int \rho_{ex}~d\cos\theta\right|$. The green line has been obtained by solving the quantum kinetic equations, focusing on  the linear regime. The orange line shows the growth predicted by the linear stability analysis for the homogeneous mode. The excellent agreement between these two independent approaches in the linear regime suggest that the inhomogeneous modes might negligibly affect the growth rate.}
\label{LSAcomp}
\end{figure}

The  impact of inhomogeneities on the growth rate can also be explored     solving the neutrino quantum kinetic equations numerically  in the linear regime and comparing the latter with the growth rate resulting from the linear stability analysis carried out considering the homogeneous mode.
Figure~\ref{LSAcomp} shows such comparison at  $r=21.3$~km for $t_{\rm{pb}}=1$~s. We can see  good agreement. 
 We have carried out this comparison for all snapshots and a range of radii for each time snapshot. To this purpose, however, we have used a self-interaction strength three orders of magnitude smaller than the one computed relying on the neutrino properties. The attenuation of the self-interaction strength has been employed  because of the technical challenges due to the fact that  the numerical simulation has to be performed for very long time to see the exponential growth for the very small growth rates typical of certain radii. In all cases, we find good agreement  between the numerical solution of the quantum kinetic equations and the linear stability analysis carried out for the homogeneous mode  (results not shown here as comparable to the ones reported in Fig.~\ref{LSAcomp}).  We note that we also find good agreement between the results of the linear stability analysis carried out considering the homogeneous mode (cf.~the red dashed line of Fig.~\ref{growth}) and the solution of the neutrino quantum kinetic equations for $t_{\rm{pb}}=0.12$~s . As shown in Fig.~\ref{growth}, for this snapshot fast instabilities are irrelevant for the homogeneous mode despite the existence of ELN crossings (cf.~green line); because of the interplay between $\omega_{\rm vac} \neq 0$ and  the ELN crossing, the system becomes unstable, with  a growth rate represented by  the blue curve. We note that the employment of the attenuated self-interaction strength may overestimate the relevance of the vacuum term; nevertheless our main conclusions would not be qualitatively affected.

%While these findings should be taken with caution due to the approximations involved, they call for dedicated work to assess the interplay of inhomogeneities with fast and collisional instabilities. To this purpose, a method to carry out the linear stability analysis in the presence of inhomogeneities and in the absence of periodic boundaries should be also worked out.

Our findings seem to suggest that, in the absence of periodic boundary conditions, there may be a unique growth rate that is of relevance and it is given by the solution of Eqs.~\ref{lin1sp} and \ref{lin2sp}. This is different from what is  considered in Ref.~\cite{Xiong:2022vsy}, when the growth rate is calculated for several  Fourier modes; in this case, each Fourier mode has its own growth rate, the largest of them is interpreted as the physically relevant one. However, such an analysis assumes self-sustaining perturbations at all possible length scales, which is possible only if periodic boundary conditions are imposed. 
While our findings should be taken with caution due to the approximations involved, they call for dedicated work to assess the interplay of inhomogeneities with fast and collisional instabilities. To this purpose, a method to carry out the linear stability analysis in the presence of inhomogeneities and in the absence of periodic boundaries should be also worked out.

\section{Discussion and outlook}
\label{sec:conclusions}
The coherent forward scattering of neutrinos onto each other is responsible for triggering flavor instabilities that could be fast (if driven by an ELN crossing for $\omega_{\rm vac} =0$) or slow (if the neutrino vacuum frequency also plays a role). In addition, more recently, it has been advanced the possibility that also the collision of neutrinos with the background medium could be responsible for triggering flavor conversion~\cite{Johns:2021qby, Johns:2022yqy, Martin:2021xyl, Lin:2022dek,Padilla-Gay:2022wck,Xiong:2022vsy,Xiong:2022zqz,Kato:2023dcw}. 

References~\cite{Shalgar:2022lvv,Shalgar:2022rjj} pointed out the possible co-existence of slow and fast instabilities in the neutrino decoupling region.  Because of the non-negligible collision rate, in this paper, we aim to investigate the interplay among slow, fast, and collisional instabilities in the SN core. To this purpose, we rely on static hydrodynamic backgrounds and thermodynamic properties extracted at post-bounce times between $0.05$ and $1$~s from a spherically symmetric hydrodynamical simulation of a SN with mass $18.6\ M_\odot$~\cite{SNarchive}. We compute the neutrino angular distributions in the absence of flavor conversion solving the neutrino kinetic equations in the decoupling region for an average energy mode. 

We then carry out the linear stability analysis for the homogeneous mode and investigate the occurrence of slow, fast, or collisional flavor instabilities. We find that, due to the ubiquitous existence of ELN crossings, the fast flavor instability dominates virtually in all cases, except for one time snapshot ($t_{\rm pb}=0.12$~s) where the slow flavor instability dominates. Under no circumstances did we find the collisional instability to be the primary contributor to the flavor instability.

Our findings are in contrast with the analysis presented in Ref.~\cite{Xiong:2022vsy}. This could be due to several reasons, such as the application in Ref.~\cite{Xiong:2022vsy} of the attenuation method~\cite{Nagakura:2022kic} to the self-interaction potential as well as an arbitrary rescaling of the collision term for energies above $90$~MeV, and an implementation of the collision term that is different from ours. This approach may have led to a fictitious growth of the collisional instabilities, which should not have occurred otherwise. 
%\sout{Note that a rapid change in the neutrino properties is also found in Ref.~\cite{Xiong:2022vsy} (cf.~their Fig.~10), which should not be the case for our model.}
 Note that the growth rates in Ref.~\cite{Xiong:2022vsy} (cf.~their Fig.~3) are strongly time dependent due to rapidly changing number densities, whereas we assume a classical steady state in which the number densities of neutrinos do not change in the linear regime.

We have ignored the effects of energy dependent collision terms that can be important in several ways as pointed out in Ref.~\cite{Kato:2023dcw}. However, the conclusions reached in Ref.~\cite{Kato:2023dcw} cannot be directly extended to our model since the analysis in Ref.~\cite{Kato:2023dcw}   does not assume an initial classical steady state as we do. A systematic study of collisional instabilities with energy dependent collision terms is left to future work. The role of inhomogeneities in the development of flavor instabilities, for which we only present preliminary findings in this paper, should also be subject of future dedicated work. 

In conclusion, collisional instabilities remain an intriguing possibility to trigger flavor  conversion; however, it is not clear whether in neutrino-dense astrophysical environments, flavor instabilities due to collisions could drive and sustain flavor conversion. Further work will be needed to analyze the development of flavor instabilities in a large set of SN models, as well as neutron star merger simulations, exhibiting a wide range of thermodynamic and neutrino properties.

\acknowledgments
We thank Thomas Janka and Evan O'Connor for insightful discussions. We are also grateful to Thomas Janka, Lucas Johns, Zewei Xiong, and Meng-Ru Wu for helpful comments on our manuscript.
This project has received support from the Villum Foundation (Project No.~13164), the Danmarks Frie Forskningsfonds (Project No.~8049-00038B), and the Deutsche Forschungsgemeinschaft through Sonderforschungbereich
SFB~1258 ``Neutrinos and Dark Matter in Astro- and
Particle Physics'' (NDM).

%\bibliography{thetaphi.bib}

\appendix
\section{Collisional terms}
\label{collisions}
\begin{figure*}
\includegraphics[width=0.95\textwidth]{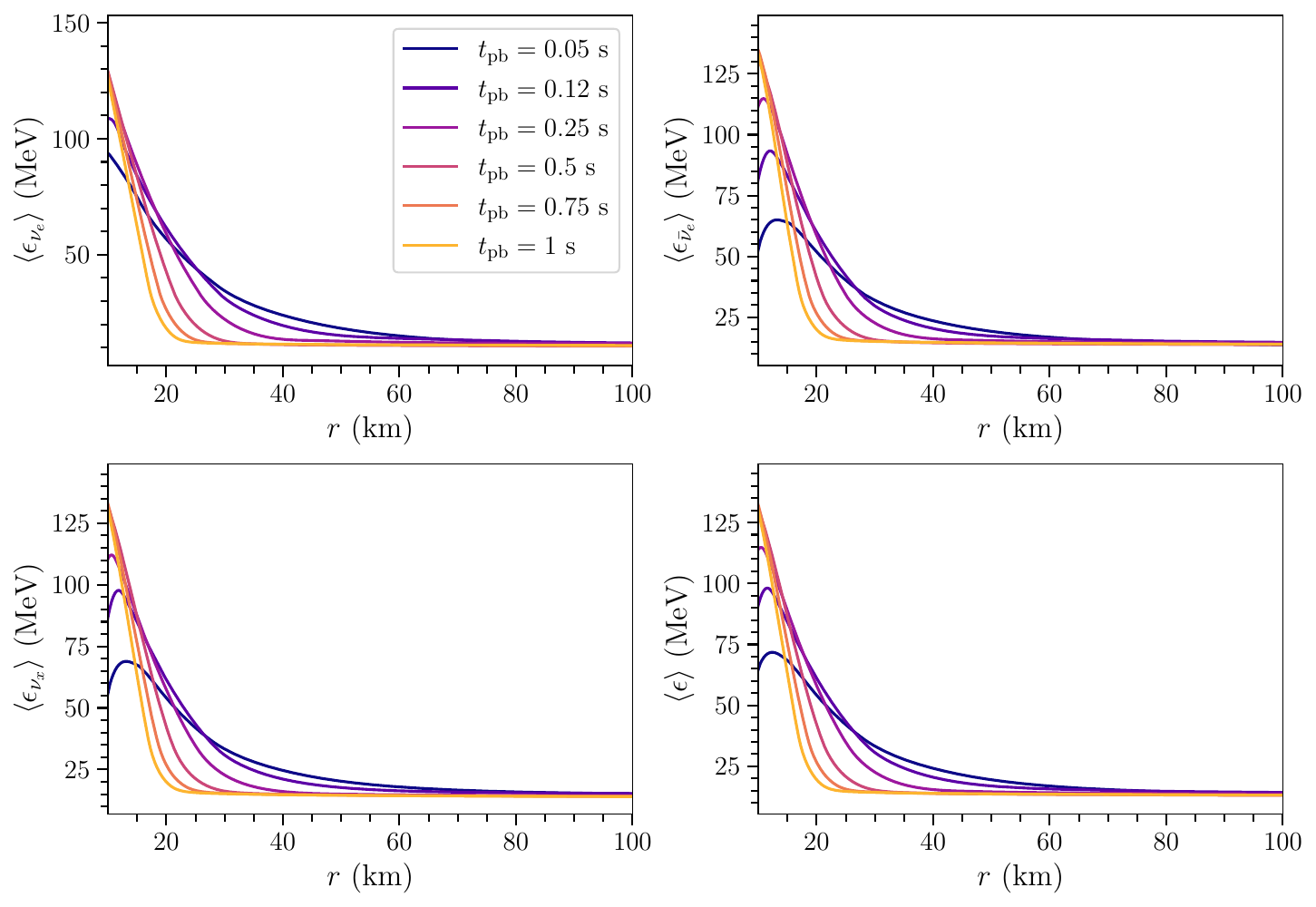}
\caption{Radial profile of the first energy moment for $\nu_e$ (top left), $\bar\nu_e$ (top right) and $\nu_x$ (bottom left) for all selected post-bounce times $t_{pb} =0.05$, $0.12$, $0.25$, $0.5$, $0.75$, and $1$~s extracted for our benchmark hydrodynamical SN simulation. The bottom right panel represents the flavor-averaged first energy moment that we use as characteristic neutrino energy in the linear stability analysis.}
\label{fig:Eave}
\end{figure*}
In the interior of a core-collapse SN, several processes contribute to the production, absorption, and interaction of neutrinos with matter. The most relevant ones are Bremsstrahlung, pair-production, beta reactions, and direction changing processes. In this appendix, we present the implementation of these terms in the neutrino kinetic equations (see Eqs.~\ref{coll1} and \ref{coll2}). Our modeling of the collisional kernel was heavily inspired by the open-source neutrino transport code NuLib~\cite{OConnor:2014sgn}.

It should be noted that there are other thermal processes occurring in the SN interior which we ignore, namely, plasmon decay, recombination, and photo-production as they are subdominant~\cite{1996ApJS}. In addition, we neglect the energy changing non-elastic neutrino-electron scattering because it is subdominant, especially for the electron type-neutrinos which determine the development of the instabilities of interest to this work. We do not consider charge current interactions on muons since our benchmark SN simulation does not include them for simplicity, despite the fact that  muons could be important~\cite{Bollig:2017lki}.

For calculating beta processes and direction changing rates, we use the flavor-dependent average energy of neutrinos extracted from our benchmark hydrodynamical SN simulation and displayed in Fig.~\ref{fig:Eave}, whereas for thermal processes (i.e.~Bremmstrahlung and pair processes) we calculate the production by integrating over the energy dependent emission rates, and calculate the total absorption rate by applying Kirchoff's law to the energy integrated production rate.

\subsection{Link between the absorption and emission rates}
The absorption and emission rates are not independent in the steady state configuration, but are related by the Kirchoff's law of radiation. As a consequence of the Kirchoff's law, the emission rate is such that the number density reaches a thermal distribution asymptotically in the absence of advection. 
The Kirchoff's law is not just valid in thermal equilibrium, but also in a steady state. Even in the low density region where the distribution of particles is far from thermal, the Kirchoff's law can be used to relate the absorption term to the emission one~\cite{thompsonthesis}: 
\begin{eqnarray}
\mathcal{L}^{\textrm{int}}[\rho_{ii}] = \eta_{\nu}(1-\rho_{ii}) - \chi_{\nu}\rho_{ii}\ , 
\label{boltzmann}
\end{eqnarray}
where $\chi_{\nu}$ is the absorption rate and $\eta_{\nu}$ is the emission rate without Pauli blocking.
The right hand side of Eq.~\ref{boltzmann} is zero in equilibrium or equivalently:
\begin{eqnarray}
\eta_{\nu} = \chi_{\nu} \frac{\rho_{ii}^{\textrm{eq}}}{1-\rho_{ii}^{\textrm{eq}}}\ ,
\label{unsimkir}
\end{eqnarray}
with $\rho_{ii}^{\textrm{eq}}$ being the Fermi-Dirac distribution of neutrinos 
Substituting, Eq.~\ref{unsimkir} in Eq.~\ref{boltzmann}, we obtain
\begin{eqnarray}
\mathcal{L}^{\textrm{int}}[\rho_{ii}] = \mathcal{C}^{ii}_{\textrm{absorb}}[\rho_{ii}^{\textrm{eq}}-\rho_{ii}]\ ,
\end{eqnarray}
where $\mathcal{C}^{ii}_{\textrm{absorb}}=\chi_{\nu}/(1-\rho_{ii}^{\textrm{eq}})$ is the absorption rate with the denominator included to account for Pauli blocking.

In the following, we discuss the emission and absorption terms. As for $\nu_{e}$ and $\bar{\nu}_{e}$, charged current interactions (beta reactions) dominate over the neutral current processes, but neutral current processes are important for heavy lepton neutrinos.

\subsection{Emission and absorption processes: Beta reactions}
Beta reactions are the dominant channels determining the number density of electron type neutrinos. The absorption and emission of $\nu_{e}$ and $\bar{\nu}_{e}$ occurs through the following interactions with free nucleons:
\begin{eqnarray}
\nu_{e} + n & \leftrightarrow & p+e^{-} \\
\bar{\nu}_{e} + p & \leftrightarrow & n+e^{+}\ .
\end{eqnarray}
The correction to the beta reaction rates due to the presence of heavier nuclei is subdominant and can be neglected for our purposes. 

The absorption cross section of electron neutrinos on neutrons is given by~\cite{Bruenn:1985en}:
\begin{eqnarray}
\sigma^{a}_{\nu_{e}n} &=& \sigma_{0} \left(\frac{1+3g_{A}^{2}}{4}\right)\left(\frac{\langle \epsilon_{\nu_{e}} \rangle +\Delta_{np}}{m_{e}c^{2}}\right)^{2} \nonumber \\
& \times & \left[1-\left(\frac{m_{e}c^{2}}{\langle \epsilon_{\nu_{e}} \rangle +\Delta_{np}}\right)^{2}\right]^{\frac{1}{2}}(1-f_{e^{-}})W_{M}\ .
\label{signue}
\end{eqnarray}
Here, $\sigma_{0}$ is the characteristic neutrino cross section, 
\begin{eqnarray}
\sigma_{0} = \frac{4 G_{\textrm{F}} \left(m_{e}c^{2}\right)^{2}}{\pi (\hbar c)^{4}} = 1.7612 \times 10^{-44} \textrm{cm}^{2}\ ,
\label{eq:sigma0}
\end{eqnarray}
with $g_{A}=-1.254$, $\Delta_{np}$ being the difference in the mass of the neutron and proton ($m_{n}-m_{p}=1.2933$ MeV), $m_{e}$ the mass of the electron, $\langle \epsilon_{\nu_e} \rangle$ the $\nu_e$ average energy displayed in Fig.~\ref{fig:Eave}, $f_{e^{-}}$ the Fermi-Dirac distribution of electrons, and $W_{M} = 1-1.01\langle \epsilon_{\nu_{e}} \rangle /m_{n}$ the weak magnetic correction~\cite{Vogel:1983hi, thompsonthesis}.

The related inverse transport mean free path of neutrinos is
\begin{eqnarray}
\mathcal{C}_{\textrm{absorb}} = \sigma^{a}_{\nu_{e}n} X_{n} \rho_B N_{A} (1-\rho_{ee}^{\textrm{eq}})^{-1}\ ,
\end{eqnarray}
where $\sigma^{a}_{\nu_{e}n}$ is given by Eq.~\ref{signue}, and $X_{n}$ and $\rho_B$ are the neutron fraction and the baryon density shown in Fig.~\ref{thermo}. 
In the expression above, we ignore the Pauli blocking of nucleons as it is subdominant in the decoupling and free streaming region~\cite{Bruenn:1985en,Raffelt:1996wa}.

A similar expression holds for $\bar{\nu}_{e}$ absorption with the exception that we ignore the Pauli blocking for the outgoing positron and with other appropriate changes following Ref.~\cite{Bruenn:1985en}:
\begin{eqnarray}
\sigma^{a}_{\bar{\nu}_{e}p} &=& \sigma_{0} \left(\frac{1+3g_{A}^{2}}{4}\right)\left(\frac{\langle \epsilon_{\bar{\nu}_{e}} \rangle -\Delta_{np}}{m_{e}c^{2}}\right)^{2} \nonumber \\
& \times & \left[1-\left(\frac{m_{e}c^{2}}{\langle \epsilon_{\bar{\nu}_{e}} \rangle -\Delta_{np}}\right)^{2}\right]^{\frac{1}{2}}W_{\bar{M}}\ ,
\end{eqnarray}
with $W_{\bar{M}} = (1-7.1\langle \epsilon_{\bar{\nu}_{e}}\rangle/m_{p})$. 
The inverse of the transport mean free path is defined as
\begin{eqnarray}
\bar{\mathcal{C}}_{\textrm{absorb}} = \sigma^{a}_{\bar{\nu}_{e}p} X_{p} \rho_B N_{A} (1-\bar{\rho}_{ee}^{\textrm{eq}})^{-1}\ .
\end{eqnarray}

\begin{figure*}
\includegraphics[width=0.99\textwidth]{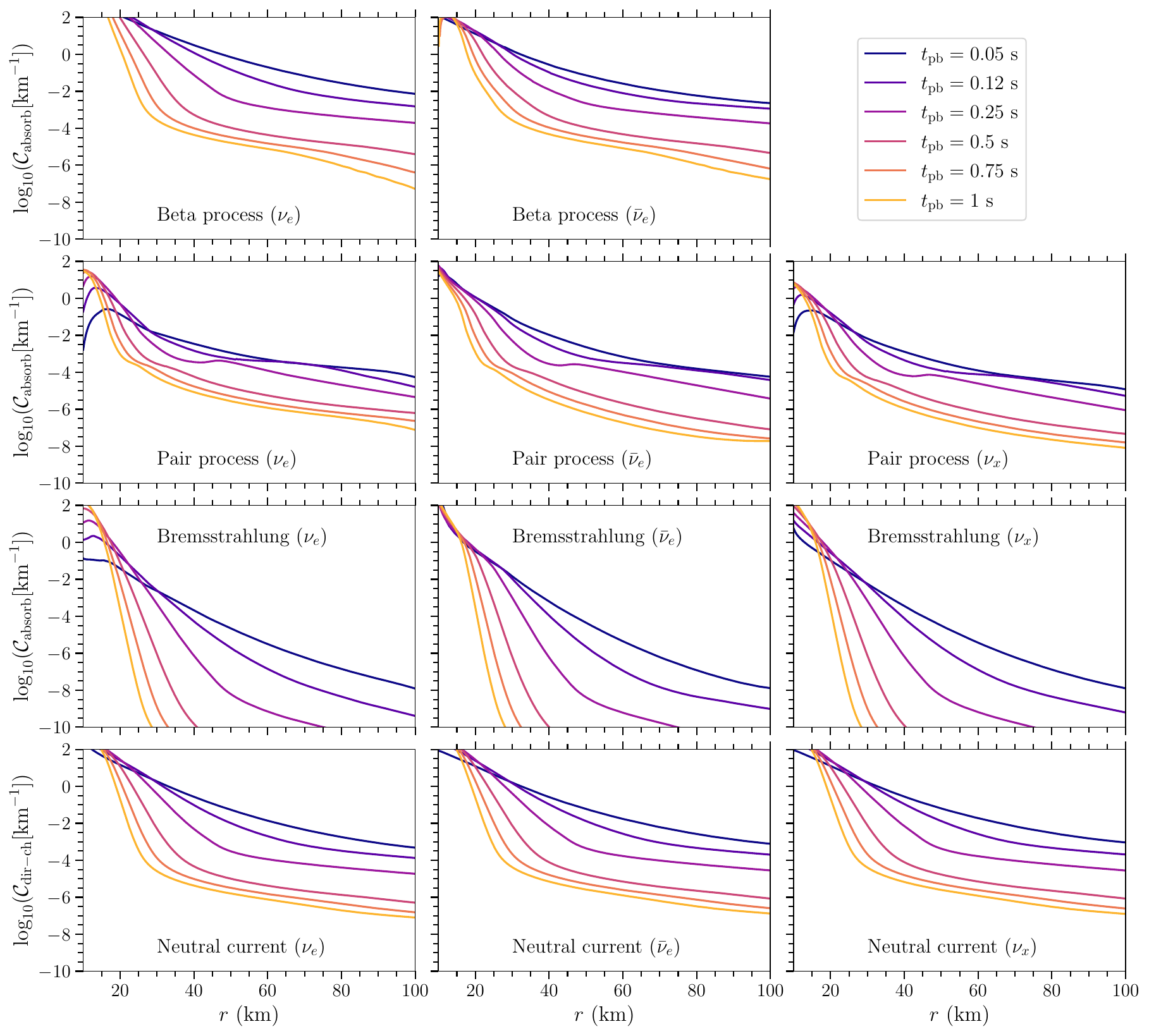}
\caption{Radial evolution of the absorption rates due to beta reactions for $\nu_{e}$ and $\bar{\nu}_{e}$ (top row; the emission rate is linked to the absorption one through the Kirchoff's law), pair production (second row), Bremsstrahlung (third row), and direction changing neutral current scattering processes (bottom row) that we compute at various post-bounce time snapshots and for $\nu_e$ (left panels), $\bar\nu_e$ (middle panels), and $\nu_x$ (right panels). The difference in the interaction rate for the different flavors is due to the different average energy of each flavor. The emission rates are related to the absorption rates by the Kirchoff's law. At small $r$, beta processes are dominant. At larger radii, direction-changing reactions start to become competitive with beta processes and are more efficient than pair and bremsstrahlung reactions, with the latter rapidly decreasing as a function of time. 
}
\label{betaabsorb}
\end{figure*}

The absorption mean free path can be used to calculate the emission rate relying on the Kirchoff's law:
\begin{eqnarray}
\mathcal{C}_{\textrm{emit}} &=& \mathcal{C}_{\textrm{absorb}} \rho_{ii}^{\textrm{eq}}\\
\bar{\mathcal{C}}_{\textrm{emit}} &=& \bar{\mathcal{C}}_{\textrm{absorb}} \bar{\rho}_{ii}^{\textrm{eq}}
\end{eqnarray}
It should be noted that the equality above holds for a steady state configuration and it is not necessary to have thermal equilibrium. Hence, we can use the Kirchoff's law in the trapped regime, decoupling regime, and the free streaming regime. 
The top panels of Fig.~\ref{betaabsorb} display the radial evolution of the absorption rates due to beta processes for electron neutrinos and antineutrinos for our selected post-bounce snapshots. 

\subsection{Emission and absorption processes: Pair production}

Electrons and positrons annihilate to produce pairs of neutrinos and antineutrinos:
\begin{eqnarray}
e^{-} + e^{+} \leftrightarrow \nu_{i}+\bar{\nu}_{i} \quad i=e, x\ .
\end{eqnarray}
The $s$-channel Feynman diagrams can create pairs of neutrinos of all flavors, along with the $t$-channel diagrams that can only produce pairs of electron type neutrinos. Note that the electron-positron pairs responsible for neutrino production are relativistic in the SN interior and, hence, the production cross section depends on the momenta of the initial pair. 

To compute the production rate of the neutrinos, the production kernel can be decomposed in Legendre polynomials as described in Refs.~\cite{Bruenn:1985en}. 
Assuming that the initial electron-positron pairs are isotropically distributed, 
we use Eq.~C63 of Ref.~\cite{Bruenn:1985en} to calculate the production rate for all neutrino species. Note that the produced neutrinos do not have an isotropic distribution. 

The emission rate can be calculated from the absorption rate. However, there are some caveats that need to be addressed. Unlike the beta processes, absorption occurs in pairs. This implies that the $\nu_{e}$ absorption rate due to pair processes is overestimated in the region where there are far more $\nu_{e}$'s than $\bar{\nu}_{e}$'s. This effect is not very important for electron type neutrinos as the beta reaction rates dominate over the pair processes. 
Hence we ignore the implications of such an approximation, but it is possible to add a correction to this overestimation of the absorption rate due to pair process~\cite{Betranhandy:2022lnp}.
The second row of Fig.~\ref{betaabsorb} shows the absorption rates integrated over the neutrino energy distribution for all three neutrino flavors for different snapshots.

\subsection{Emission and absorption processes: Bremsstrahlung}

The reaction $n+n \leftrightarrow n+n +\nu+\bar{\nu}$ is a subdominant thermal production process responsible for the creation of all flavors of neutrinos; as a consequence, it has virtually no impact on the neutrino distributions of all flavors, nevertheless, we include it in our calculations. Even in the case of heavy lepton neutrinos, the bremsstrahlung channel can be ignored in most cases, as visible from Fig.~\ref{betaabsorb}. 

For estimating the production and absorption rates of neutrinos through Bremsstrahlung we use the following approximate formula~\cite{Burrows:2004vq}:
\begin{eqnarray}
\frac{dQ^{\prime}_{\nu}}{d\epsilon_{\nu}} &=& \frac{0.234 Q_{nb}}{T} \left(\frac{\epsilon_{\nu}}{T}\right)^{2.4} e^{-1.1\epsilon_{\nu}/T}\ ,\\
\label{dQdE}
Q_{nb} &=& 2.0778\times 10^{30} \zeta X_{n}^{2} \rho_{B, 14}^{2} \left(\frac{T}{\textrm{MeV}}\right)^{5.5}\ ,
\label{Qnb}
\end{eqnarray}
with $T$ being the temperature, $\rho_{B, 14}$ the baryon density in units of $10^{14}$~g/cm$^3$, $X_{n}$ the neutron fraction (see Fig.~\ref{thermo}), and $\zeta = 0.5$ the correction factor~\footnote{It should be noted we follow Ref.~\cite{OConnor:2014sgn} and Eq.~\ref{Qnb} differs from Eq.~142 of Ref.~\cite{Burrows:2004vq} because of an error in that paper~\cite{O'Connor}. 
} 
Equation~\ref{dQdE} is accurate up to $3\%$~\cite{thompsonthesis}. 
 The emission term associated with this process is obtained by integrating the emissivity over the energy range,
\begin{eqnarray}
\mathcal{C}_{\textrm{emit}} = \int d\epsilon_{\nu} \frac{d Q^{\prime}_{\nu}}{d\epsilon_{\nu}}\ .
\end{eqnarray}
The absorption term is related to the emission term through the Kirchoff's law, 
\begin{eqnarray}
\mathcal{C}_{\textrm{absorb}} = \frac{\mathcal{C}_{\textrm{emit}}}{n_{i}^{\textrm{eq}}}
\end{eqnarray}
which is shown in the third row of Fig.~\ref{betaabsorb}.

\subsection{Direction changing processes: Neutral current interactions}

The momentum changing interaction of neutrinos with matter can be due to neutral current interactions as well as charged current interactions. As for the latter, only the electron flavors ($\nu_{e}$ and $\bar{\nu}_{e}$) can participate in charged current interactions with electrons; yet, this momentum changing process is subdominant with respect to the neutral current scattering of neutrinos with nucleons. Therefore, in the following, we focus on neutral current processes.

We calculate the momentum changing cross section in the single energy approximation (relying on the flavor-dependent neutrino average energy shown in Fig.~\ref{fig:Eave}), so the momentum changing cross section can be viewed as a direction changing one. 
The direction changing cross section also depends on the matter composition. 
However, for our purposes, the contribution of nuclei (and related nuclear matrix elements) can be ignored. Hence we focus on the interaction of neutrinos with protons and neutrons. We also ignore the neutral current interactions of neutrinos with electrons 
since they are subdominant with respect to the interaction with nucleons.

As the interaction of neutrinos with nucleons is independent on the neutrino flavor, the transport cross section with neutrons and protons is given by~\cite{thompsonthesis}:
\begin{eqnarray}
\sigma_{n}^{\textrm{tr}} &=& \frac{\sigma_{0}}{4}\left(\frac{\langle \epsilon_{\nu}\rangle}{m_{e}c^{2}}\right)^{2}\left(\frac{1+5 g_{A}^{2}}{6}\right)\ , \\
\sigma_{p}^{\textrm{tr}} &=& \frac{\sigma_{0}}{6}\left(\frac{\langle \epsilon_{\nu} \rangle}{m_{e}c^{2}}\right)^{2}\left[(C_{V}^{\prime}-1)^{2}+5g_{A}(C_{A}^{\prime}-1)^{2}\right]\ ,\nonumber \\
\end{eqnarray}
where $\sigma_{0}$ is defined as in Eq.~\ref{eq:sigma0}, $g_{A} = 1.254$, $C_{V}^{\prime}=1/2+2 \sin^{2}\theta_{W}$ (with $\theta_{W}$ being the Weinberg angle), and $C_{A}^{\prime}=1/2$. 
 The direction changing collision term $C_{\textrm{dir-ch}}$ as follows:
\begin{eqnarray}
\mathcal{C}_{\textrm{dir-ch}} = \rho_{B}N_{A} (X_{n} \sigma_{n}^{\textrm{tr}} + X_{p} \sigma_{p}^{\textrm{tr}}) \ .
\end{eqnarray}
It should be noted that the direction changing neutral current interactions are not isotropic, but backward peaked. The neutrino interactions with protons have an anisotropy of about 10\% ($\mathcal{C}_{\textrm{ani}}=0.1~\mathcal{C}_{\textrm{dir-ch}}$) whereas the interactions with neutrons have an anisotropy of about 33\% ($\mathcal{C}_{\textrm{ani}}=0.33~\mathcal{C}_{\textrm{dir-ch}}$). This is taken into account through $\mathcal{C}_{\rm{ani}}$, which is the weighted average of the two anisotropies in Eqs.~\ref{coll1} and \ref{coll2}. 
The bottom panels of Fig.~\ref{betaabsorb} illustrate the dependence of the direction-chancing processes on the radius; note that the flavor dependence of the neutral current interaction rate in Fig.~\ref{betaabsorb} is due to the fact that the average energy of neutrinos depends on the neutrino flavor. 
%%%%%%%%%%%%%%%%%%%%%%%%%%%%%%%%%%%%%%%%%%%%%%%%%%%%%%%%%%%%%%%%%%%%%%%%%%%%%%%
%%%%%%%%%%%%%%%%%%%%%%%%%%%%%%%%%%%%%%%%%%%%%%%%%%%%%%%%%%%%%%%%%%%%%%%%%%%%%%%

\section{Comparison of the neutrino number densities with the output of our benchmark supernova model}
\label{sec:comparison}
\begin{figure*}
\includegraphics[width=0.99\textwidth]{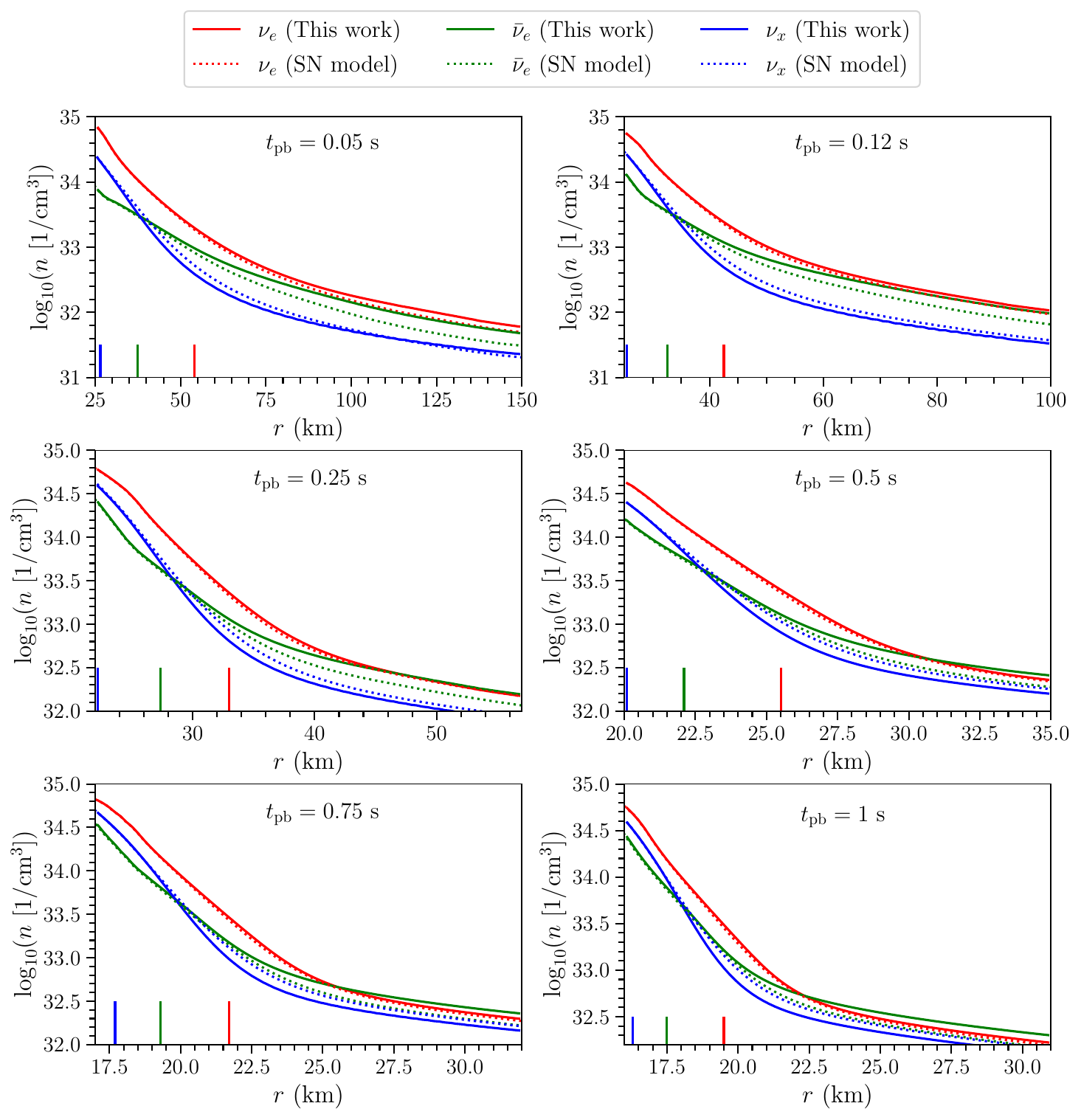}
\caption{Radial evolution of the neutrino number densities computed in this work (solid lines) of $\nu_{e}$ (in red), $\bar{\nu}_{e}$ (in green), and $\nu_{x}$ (in blue) along with  the ones extracted from our benchmark SN simulation (denoted by ``SN model'' in the legend) for selected post-bounce time snapshots. The radius at which the number density of each neutrino flavor deviates from the Fermi-Dirac one corresponds to the radius of neutrino decoupling marked by a vertical line. 
While involving some approximations, our method allows us to reproduce the local neutrino density which is in general agreement the output of the hydrodynamical simulation. Note that the radial range is not the same for all time snapshots. 
}
\label{numden1}
\end{figure*}

Relying on the hydrodynamical and thermodynamic quantities, local neutrino number density, and first energy moment extracted at selected post-bounce times from our benchmark SN simulation, we  compute the neutrino angular distributions, following the modeling of the collision term outlined in Appendix~\ref{collisions}. In this appendix, we compare the flavor-dependent and angle-integrated neutrino number densities that we obtain from post-processing the output of our $18.6\ M_\odot$ SN model~\cite{SNarchive} with the neutrino number density extracted from the hydrodynamical simulation to assess the degree of accuracy of our collision term. The comparison is shown in Fig.~\ref{numden1}. 
Despite the good agreement between the neutrino number densities that we compute in post-processing and the outcome of hydrodynamical simulations, we stress that the purpose of this work is not to reproduce the neutrino properties from hydrodynamical simulations, but rather rely on angular distributions derived from first principles to investigate the relative importance of collisional, fast, and slow flavor instabilities. 

In the SN core, neutrinos are continuously produced and absorbed at the same time (see Appendix~\ref{collisions}). These two processes eventually lead to thermal equilibration. However, the baryon density falls rapidly as a function of radius (see Fig.~\ref{thermo}), leading neutrinos to go out of thermal equilibrium and stream freely. The radius of decoupling at which this happens depends on the magnitude of the production and absorption rates compared to the direction changing term.
As shown in Fig.~\ref{numden1}, this implies that the neutrino number density coincides with the number density obtained by integrating Fermi-Dirac distribution over energy in the dense region, and it deviates from the energy-integrated Fermi-Dirac number density at larger radii where the matter density is lower, and the angular distribution becomes forward peaked. In order to assess the location of the neutrino decoupling surfaces in Fig.~\ref{numden}, we compare the local neutrino number density with the one obtained from the energy-integrated Fermi-Dirac neutrino density  and define the neutrino decoupling radius as the one where the neutrino number density deviates from the Fermi-Dirac one more than $1\%$ (see vertical lines in Fig.~\ref{numden1}).

\begin{figure}
\includegraphics[width=0.49\textwidth]{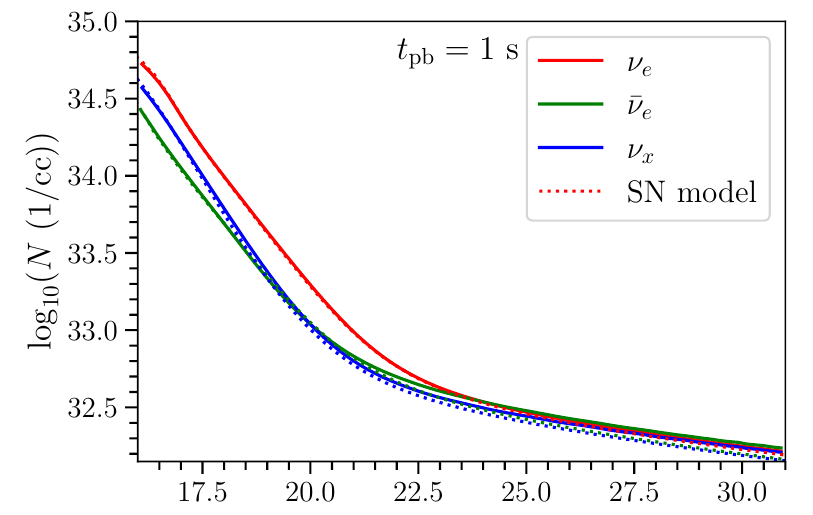}
\includegraphics[width=0.49\textwidth]{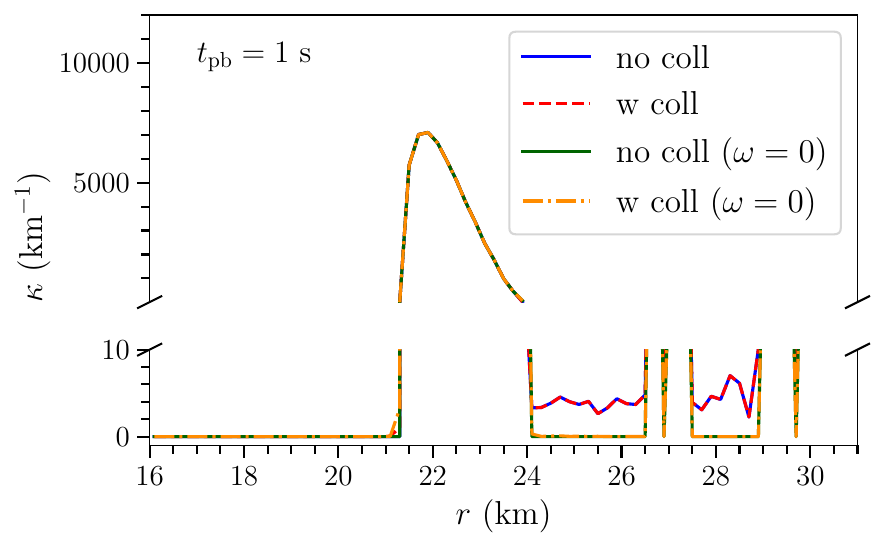}
\caption{{\it Top:} Comparison between the neutrino number densities obtained employing  multi-energy dependence in the collisional rate (solid lines) and the ones coming from our benchmark SN simulation  (dotted lines) for $t_{\textrm{pb}}=1$~s. The already small descrepancy between the number densities seen in Fig.~\ref{numden1} is further alleviated by the multi-energy treatment. {\it Bottom:} Growth rate of the flavor  instability obtained employing  multi-energy dependence in the collisional rate. The qualitative feature of collisional instability not being important remains unchanged by the single energy approximation.}
\label{mulenerrgy}
\end{figure}

 It should be noted that in Fig.~\ref{numden1} the $\bar{\nu}_{e}$ number density is slightly larger than the one from the ``SN model" because of the single energy approximation employed in this work. We demonstrate this by performing an energy-dependent computation   and report the results in Fig.~\ref{mulenerrgy} for the snapshot $t_{\textrm{pb}}$=1~s. The top panel of  Fig.~\ref{mulenerrgy} shows that the discrepancy between the number densities obtained in this work and the ones from the Garching simulation is alleviated almost completely. The remaining negligible residual difference is due to the fact that our implementation of the collisional term is not identical to the one employed in the Garching simulation. 

The bottom panel of Fig.~\ref{mulenerrgy} shows that the  growth rate obtained performing a multi-energy computation remains qualitatively unchanged (cf.~bottom right panel of Fig.~\ref{growth}).  It should be noted, however, that there are some quantitative differences between the single energy and multi-energy calculations: in the multi-energy simulation the maximum growth rate is slightly smaller and the region of instability is shifted to slightly larger radii. However, this does not affect our main conclusions. Figure~\ref{mulenerrgy}, hence,  demonstrates that the single-energy approximation is justified in our case.

We note that the  green dotted line (i.e., $\bar{\nu}_e$)  is below the  red one (i.e.~$\nu_e$) in Fig.~\ref{numden1}; on the other hand, the solid and red lines cross at $\sim 22.6$~km and the $\bar{\nu}_e$ number density is larger than the $\nu_e$'s one  at larger radii. Such crossing between the $\nu_e$ and $\bar\nu_e$ number densities is also present in our multi-energy simulations in Fig.~\ref{mulenerrgy}, although much less prominent, and it might give rise to  differences in the development of ELN crossings. 
We conclude that  the single energy approximation slightly overestimates the number density $\bar{\nu}_e$ facilitating the presence of a sharper ELN crossing. However, although shallower, an ELN crossing is also expected in the multi-energy calculation (cf.~Fig.~\ref{mulenerrgy}). Thus implies that our single-energy treatment does not affects the relative importance of fast vs.~collisional instabilities--the latter being dominant only in the absence of ELN crossings. 
In fact, a comparison between the bottom panel of Fig.~\ref{mulenerrgy}  and the bottom right panel of Fig.~\ref{growth} shows that the growth rate for fast instabilities already develops at smaller radii (the ELN crossing developing at $20.5$~km for the single-energy simulation vs.~$21.1$~km for the multi-energy simulations) and is qualitatively unchanged by this feature.  

%%%%%%%%%%%%%%%%%%%%%%%%%%%%%%%%%%%%%%%%%%%%%%%%%%%%%%%%%%%%%%%%%%%%%%%%%%%%%%%
%%%%%%%%%%%%%%%%%%%%%%%%%%%%%%%%%%%%%%%%%%%%%%%%%%%%%%%%%%%%%%%%%%%%%%%%%%%%%%%
\bibliography{collinstab.bib}
%%%%%%%%%%%%%%%%%%%%%%%%%%%%%%%%%%%%%%%%%%%%%%%%%%%%%%%%%%%%%%%%%%%%%%%%%%%%%%%
%%%%%%%%%%%%%%%%%%%%%%%%%%%%%%%%%%%%%%%%%%%%%%%%%%%%%%%%%%%%%%%%%%%%%%%%%%%%%%%
\end{document}